\documentclass[aps,prb,times,twocolumn,amsmath,amssymb,superscriptaddress,floatfix,showpacs]{revtex4}

\usepackage{color}

\usepackage{graphicx}
\usepackage{subfigure}
\usepackage{bbold}
\usepackage{verbatim}
\usepackage{float}
\usepackage{enumerate}
\usepackage{amsfonts}
\usepackage{multirow}
\usepackage[colorlinks,bookmarks=false,citecolor=blue,linkcolor=red,urlcolor=blue]{hyperref}

\newcommand{\dg}{^\dagger}
\newcommand{\pdg}{^{\phantom\dagger}}

\begin{document}


\title{
Exploring the spin-orbital ground state of Ba$_3$CuSb$_2$O$_9$
}


\author{Andrew Smerald}

\affiliation{Institut de Th{\'e}orie des Ph{\'e}nom\`{e}nes Physiques, Ecole Polytechnique F{\'e}d{\'e}rale de Lausanne (EPFL), CH-1015 Lausanne, Switzerland}

\author{Fr{\'e}d{\'e}ric Mila}
\affiliation{Institut de Th{\'e}orie des Ph{\'e}nom\`{e}nes Physiques, Ecole Polytechnique F{\'e}d{\'e}rale de Lausanne (EPFL), CH-1015 Lausanne, Switzerland}


\date{\today}


\begin{abstract}  
Motivated by the absence of both spin freezing and a cooperative Jahn-Teller effect at the lowest measured temperatures, we study the ground state of Ba$_3$CuSb$_2$O$_9$.
We solve a general spin-orbital model on both the honeycomb and the decorated honeycomb lattice, revealing rich phase diagrams.
The spin-orbital model on the honeycomb lattice contains an {\sf SU(4)} point, where previous studies have shown the existence of a spin-orbital liquid with algebraically decaying correlations.
For realistic parameters on the decorated honeycomb lattice, we find a phase that consists of clusters of nearest-neighbour spin singlets, which can be understood in terms of dimer coverings of an emergent square lattice.
While the experimental situation is complicated by structural disorder, we show qualitative agreement between our theory and a range of experiments.
\end{abstract}


\pacs{
75.10.Kt,	
75.25.Dk,	
75.47.Lx	
}
\maketitle


\section{Introduction}


The interplay of spin and orbital degrees of freedom often leads to frustration, and can give rise to unusual quantum ground states.
For example, in the triangular lattice material LiNiO$_2$, it is proposed that a Ni$^{3+}$ orbital degeneracy drives a spin-orbital resonating valence bond state\cite{vernay04,vernay06}.
Also, it has been suggested that the spinel material FeSc$_2$S$_4$ realises a disordered spin-orbital singlet ground state, with a highly suppressed gap due to proximity to a quantum critical point\cite{chen09-PRL,chen09-PRB}.

Here we concentrate on the spin-orbital ground state of the honeycomb lattice material Ba$_3$CuSb$_2$O$_9$, which has recently garnered much interest\cite{nakatsuji12,quilliam12,zhou11,ishiguro13,katayama14,shanavas14,nasu13}.
One reason for this interest arises from the theoretical finding that a spin-orbital model on the honeycomb lattice, tuned to a high symmetry {\sf SU(4)} point, realises a gapless spin-orbital liquid with algebraically decaying correlation length \cite{corboz12}.

Ba$_3$CuSb$_2$O$_9$ contains octahedrally coordinated Cu$^{2+}$ ions, with $3d^9$ configuration.
Naively, the hole associated with each Cu$^{2+}$ ion has a fourfold degeneracy: a twofold degeneracy due to the spin-1/2 degree of freedom and a two-fold degeneracy of the $d^{\sf x^2-y^2}$ and $d^{\sf 3z^2-r^2}$ $e_g$ orbitals. 
In general this degeneracy would be lifted at low temperature by long-range ordering of both the spin and orbital degrees of freedom, which would drive a cooperative Jahn-Teller distortion of the oxygen octahedra. 
However, in the case of Ba$_3$CuSb$_2$O$_9$, muon spin relaxation experiments show an absence of spin freezing down to 20mK [\onlinecite{quilliam12}], while x-ray diffraction measurements see no evidence for a cooperative Jahn-Teller effect at temperatures as low as 12K [\onlinecite{nakatsuji12,katayama14}].
This has lead to suggestions that a spin-orbital liquid state is realised\cite{nakatsuji12,nasu13,ishiguro13,katayama14}.

\begin{figure}[ht]
\centering
\includegraphics[width=0.45\textwidth]{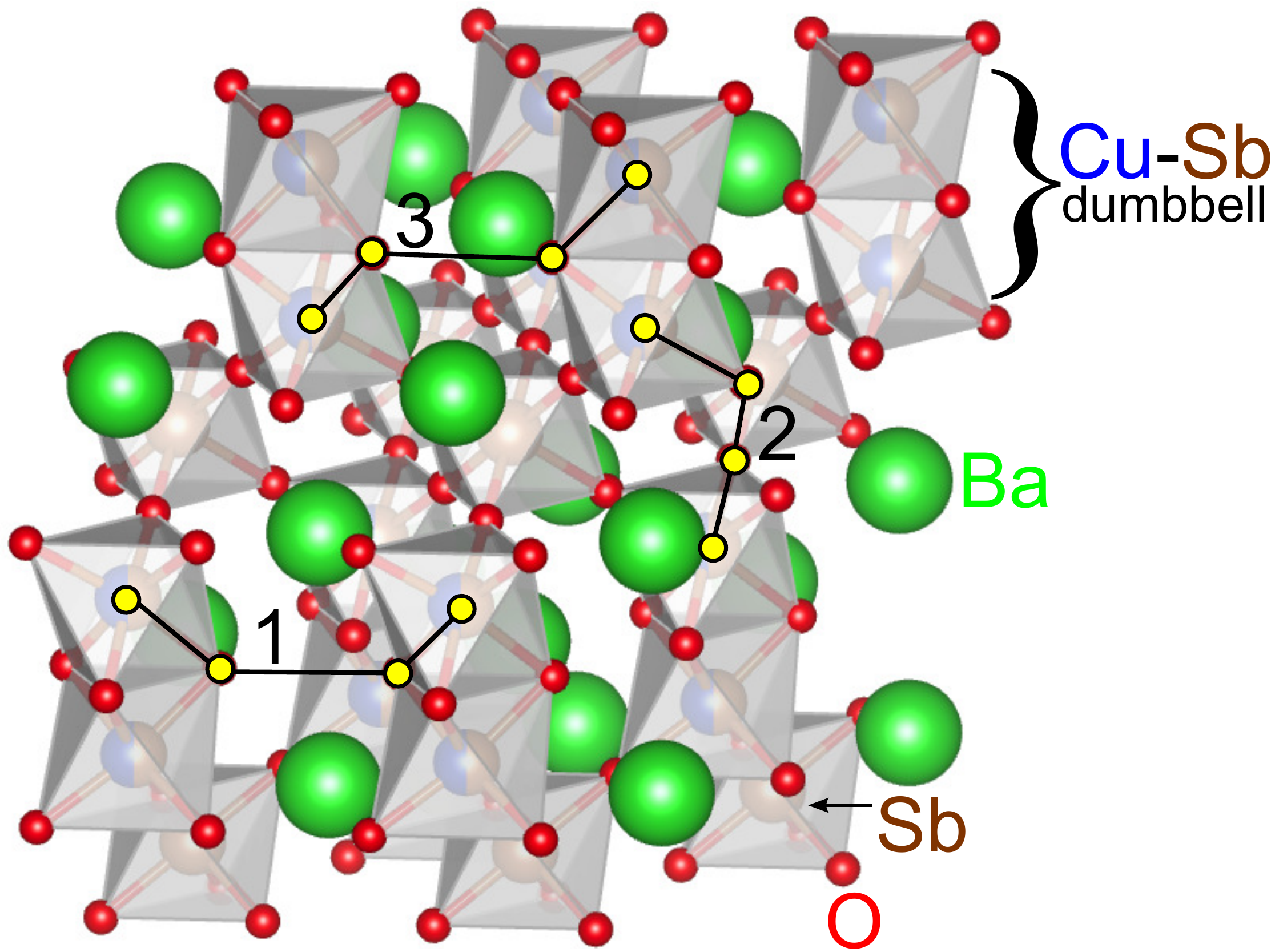}
\caption{\footnotesize{
The crystal structure of Ba$_3$CuSb$_2$O$_9$ [\onlinecite{nakatsuji12}]. 
Cu$^{2+}$-Sb$^{5+}$ dumbbells are surrounded by an oxygen bioctahedra, and form a triangular lattice.
An electric-dipole interaction between the dumbbells favours antiparallel nearest-neighbour alignment, and this results in a short-range ordered honeycomb lattice of Cu$^{2+}$ ions.
Cu$^{2+}$ has the electron configuration $3d^9$, and, therefore, there is on average one hole per Cu site.
The dominant interaction is superexchange via Cu-O-O-Cu pathways (shown by yellow dots).
Due to the bonding angles, the superexchange interaction is comparable on paths 1 and 2, but considerably weaker on path 3.
}}
\label{fig:Ba3CuSb2O9-structure}
\end{figure}

The crystal structure of Ba$_3$CuSb$_2$O$_9$ [\onlinecite{nakatsuji12}] is illustrated in Fig.~\ref{fig:Ba3CuSb2O9-structure}.
The important subunit is the Cu$^{2+}$-Sb$^{5+}$ dumbbell, which is surrounded by an O$^{2-}$ bioctahedra.
The bioctahedra have $C_{3v}$ symmetry, with the $C_3$ rotation axis parallel to the Cu-Sb bond.
Crucially, the group $C_{3v}$ has a 2-dimensional irreducible representation, which corresponds to a degeneracy between the $d^{\sf x^2-y^2}$ and $d^{\sf 3z^2-r^2}$ $e_g$ orbitals.
The Cu$^{2+}$-Sb$^{5+}$ dumbbells form a triangular lattice, and each dumbbell can be orientated with either the Cu$^{2+}$ above the Sb$^{5+}$ or vice versa.
Thus at each lattice site there is an Ising degree of freedom.
The electric dipole interaction between the dumbbells favours antiparallel nearest-neighbour alignment, and as a result one finds a short-range ordered honeycomb lattice of Cu$^{2+}$ ions\cite{nakatsuji12}.

The four-fold per-site degeneracy of the Cu$^{2+}$ hole can be lifted by either the electron exchange interaction, the electron-lattice interaction or a combination of the two.
While the spin degeneracy is clearly lifted by electron exchange, the lifting of the orbital degeneracy is more subtle.
Density functional calculations show that a Jahn-Teller distortion of the oxygen octahedra is driven by electon-lattice coupling and the consequent elastic distortion of the lattice\cite{shanavas14}.
However, at the elastic level, this effect selects a magnitude for the Jahn-Teller distortion, but not an orientation.
Thus the orbital degeneracy remains.
Lifting of this orbital degeneracy can occur either via the electron exchange interaction, or via anharmonic terms in the lattice distortion potential. 
A rough comparison of the energy scales of these two interactions gives $\sim$20meV for the exchange interaction \cite{shanavas14} and $\sim$2meV for the barrier between orbital minima due to the anharmonic lattice potential \cite{garcia10}.

As a consequence, we focus on the role of the electronic exchange interaction.
Due to the large distance between Cu ions, this is expected to be dominated by superexchange along Cu-O-O-Cu paths, and a representative selection of these paths are shown in Fig.~\ref{fig:Ba3CuSb2O9-structure}.
Analysis of the bonding angles suggests that the most important paths are within Cu planes (path 1 in Fig.~\ref{fig:Ba3CuSb2O9-structure}) and between Cu ions in neighbouring bilayers (path 2 in Fig.~\ref{fig:Ba3CuSb2O9-structure}).
Superexchange interactions between Cu atoms in different planes of the same bilayer (path 3 in Fig.~\ref{fig:Ba3CuSb2O9-structure}) are expected to be considerably weaker\cite{nakatsuji12}.

The primary focus of this article is to study the ground state of Ba$_3$CuSb$_2$O$_9$, starting from a microscopic exchange Hamiltonian.
A secondary focus is to map out the phase diagram of a realistic spin-orbital model on the honeycomb lattice.
We will find that for a sizeable region of parameter space, this supports an {\sf SU(4)} spin-orbital liquid phase.
We emphasise that we do not think this {\sf SU(4)} liquid is relevant to Ba$_3$CuSb$_2$O$_9$. 
However, it is clearly a very interesting phase, and the fact that it covers a relatively large area of parameter space lends hope to the idea that it may be realised if other honeycomb lattice compounds with active spin and orbital degrees of freedom can be synthesised.

The remainder of the paper is structured as follows.
In Section~\ref{sec:honeycomb} we consider a single plane of Cu ions, arranged on a honeycomb lattice (i.e. superexchange occurs only along path 1 in Fig.~\ref{fig:Ba3CuSb2O9-structure}).
By diagonalising small clusters, both exactly and within a mean field approximation, we map out the phase diagram as a function of the microscopic parameters.
In Section~\ref{sec:dechoneycomb} we also include the superexchange path between Cu sites in different bilayers (path 2 in Fig.~\ref{fig:Ba3CuSb2O9-structure}), and thus consider a decorated honeycomb lattice.
For realistic parameters this completely changes the ground state phase diagram, in comparison to the honeycomb case.
Finally in Section~\ref{sec:discussion} we discuss the experimental situation and consider at a qualitative level the role of structural disorder.


\section{Ground state of spin-orbital model on the honeycomb lattice}
\label{sec:honeycomb}


In this section we consider a honeycomb lattice of Cu ions.
The lattice symmetries are used to construct a microscopic exchange model for the hole degree of freedom, and from this a spin-orbital Hamiltonian is derived in second order perturbation theory.
The ground state phase diagram is calculated for small clusters of Cu sites using both exact diagonalisation and mean-field decoupling of the spin and orbital degrees of freedom.


\subsection{Microscopic Hamiltonian}


First we construct a two band Hubbard model for a honeycomb lattice of Cu sites.
Each Cu can accomodate up to four holes, labelled by the spin $\{ \uparrow, \downarrow \}$ and orbital $\{a=d^{\sf 3z^2-r^2}, b=d^{\sf x^2-y^2} \}$ quantum numbers.
The Hamiltonian is given by, 
\begin{align}
\mathcal{H}^{\sf Hub} = \mathcal{H}^{\sf hop} + \mathcal{H}^{\sf coul} , 
\label{eq:Hhub}
\end{align}
where $ \mathcal{H}^{\sf hop} $ describes the hopping of holes between neighbouring sites and $\mathcal{H}^{\sf coul} $ is an on-site Coulomb interaction.

\begin{figure*}[ht]
\centering
\includegraphics[width=0.9\textwidth]{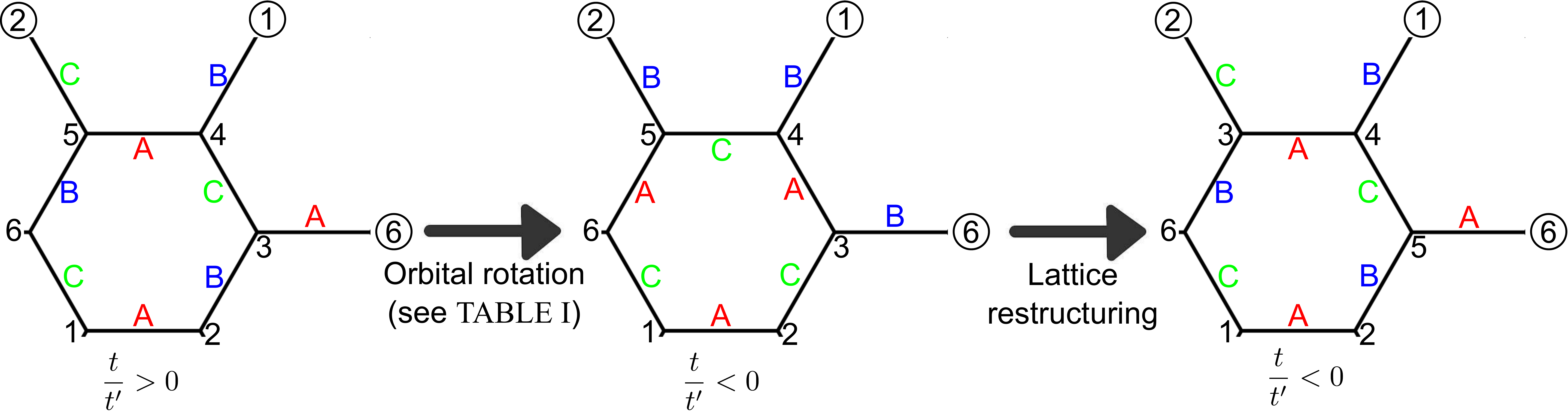}
\caption{\footnotesize{
Mapping $t/t^\prime \to -t/t^\prime$, illustrated on a 6-site cluster with periodic boundary conditions.
Bond labels {\sf A}, {\sf B} and {\sf C} correspond to different ${\bf n}_{ij}$ vectors [see Eq.~(\ref{eq:nvectors})].
Sites are numbered $1\dots 6$, and are circled to show the periodic boundary conditions.
The $t/t^\prime \to -t/t^\prime$ mapping proceeds in two steps.
First an on-site orbital rotation is performed, given in Table~\ref{tab:mapping}, and secondly a lattice restructuring, which involves swapping sites 3 and 5.
An equivalent mapping can be made on the 18-site cluster.
}}
\label{fig:mapping}
\end{figure*}

Bonds on the honeycomb lattice are labelled {\sf A}, {\sf B} and {\sf C}, as shown in Fig.~\ref{fig:mapping}.
The hopping Hamiltonian on the {\sf A} bonds is particularly simple and given by,
\begin{align}
\mathcal{H}^{\sf hop}_{ij,{\sf A}} = 
-t \sum_{\sigma=\uparrow, \downarrow} c\dg_{i,a,\sigma} c\pdg_{j,a,\sigma}
-t^\prime \sum_{\sigma=\uparrow, \downarrow} c\dg_{i,b,\sigma} c\pdg_{j,b,\sigma}
+\mathrm{H.c.},
\label{eq:Hhop}
\end{align}
where the operator $c\dg_{i,a,\sigma}$ creates a hole on the site $i$ with spin $\sigma$ and orbital $a$, and $t$ and $t^\prime$ parametrise the hopping amplitudes.
The absence of inter-orbital hopping is due to the mirror symmetry of the honeycomb lattice, under which,
\begin{align}
 c\dg_{i,a,\sigma}  \to  c\dg_{i,a,\sigma}, \quad
 c\dg_{i,b,\sigma}  \to  -c\dg_{i,b,\sigma} .
 \label{eq:mirsymtransformation}
\end{align}
Implicitly, we have chosen the orbital $z$ axis to be perpendicular to the {\sf A} bond.
The hopping amplitudes $t$ and $t^\prime$ are expected to primarily describe superexchange via Cu-O-O-Cu paths, but also include all other exhange processes between neighbouring Cu ions.

Hopping along {\sf B} and {\sf C} bonds follows from making a $\mp 2\pi/3$ rotation around the $C_3$ axis of the bioctahedra (see Fig.~\ref{fig:mapping} for bond labelling).
Under such a transformation the hole creation operators are transformed according to,
\begin{align}
 c\dg_{i,a,\sigma}  &\to  -\frac{1}{2}c\dg_{i,a,\sigma} \mp \frac{\sqrt{3}}{2}c\dg_{i,b,\sigma}, \nonumber \\
 c\dg_{i,b,\sigma}  &\to  \pm\frac{\sqrt{3}}{2}c\dg_{i,a,\sigma} - \frac{1}{2}c\dg_{i,b,\sigma} .
 \label{eq:crotation}
\end{align}
The terms generated by this transformation include intra- and inter-orbital hopping.

The on-site Coulomb interaction is described by,
\begin{align}
\mathcal{H}^{\sf coul}_i &= 
\frac{\tilde{U}}{2} n_i^2 
-J_{\sf H}\left( {\bf S}_{i,{\sf a}} \cdot {\bf S}_{i,{\sf b}} +\frac{3}{4} n_{i,{\sf a}} n_{i,{\sf b}}  \right) \nonumber \\
&+ J_{\sf p} (c\dg_{i,{\sf a},\uparrow}c\dg_{i,{\sf a},\downarrow} + c\dg_{i,{\sf b},\uparrow}c\dg_{i,{\sf b},\downarrow})  (c\pdg_{i,{\sf a},\downarrow}c\pdg_{i,{\sf a},\uparrow} + c\pdg_{i,{\sf b},\downarrow}c\pdg_{i,{\sf b},\uparrow}),
\label{eq:Hcoul}
\end{align}
where $\tilde{U}$ is the usual on-site repulsion, $J_{\sf H}$ describes the Hund's rule coupling, $J_{\sf p}$ is a pair hopping term, 
\mbox{${\bf S}_{i,{\sf a}}=(S^{\sf x}_{i,{\sf a}},S^{\sf y}_{i,{\sf a}},S^{\sf z}_{i,{\sf a}})$},
\mbox{$S^{\sf x}_{i,{\sf a}} = 1/2(c\dg_{i,{\sf a},\uparrow} c\pdg_{i,{\sf a},\downarrow}+c\dg_{i,{\sf a},\downarrow} c\pdg_{i,{\sf a},\uparrow})$},
\mbox{$S^{\sf y}_{i,{\sf a}} = -i/2(c\dg_{i,{\sf a},\uparrow} c\pdg_{i,{\sf a},\downarrow}-c\dg_{i,{\sf a},\downarrow} c\pdg_{i,{\sf a},\uparrow})$}
\mbox{$S^{\sf z}_{i,{\sf a}} = 1/2(c\dg_{i,{\sf a},\uparrow} c\pdg_{i,{\sf a},\uparrow}-c\dg_{i,{\sf a},\downarrow} c\pdg_{i,{\sf a},\downarrow})$},
 \mbox{$n_{i,{\sf a}} = c\dg_{i,{\sf a},\uparrow} c\pdg_{i,{\sf a},\uparrow} +c\dg_{i,{\sf a},\downarrow} c\pdg_{i,{\sf a},\downarrow} $} 
 and \mbox{$n_i = n_{i,{\sf a}} + n_{i,{\sf b}}$}.
Since the Cu environment is approximately cubic, we set \mbox{$2J_{\sf p}=J_{\sf H}$} [\onlinecite{castellani78,oles00}].

%


\subsection{Effective spin-orbital Hamiltonian}


In the limit $t,t^\prime \ll \tilde{U}$ an effective Kugel-Khomskii Hamiltonian can be derived, significantly reducing the Hilbert space\cite{kugel73,vernay04}.
This is written in terms of a spin-1/2 degree of freedom ${\bf S}$ and a pseudospin-1/2 orbital degree of freedom ${\bf T}$, with components,
\begin{align}
S^{\sf x}_{i} &= \frac{1}{2} \sum_{m={\sf a,b}} (c\dg_{i,{\sf m},\uparrow} c\pdg_{i,{\sf m},\downarrow}+c\dg_{i,{\sf m},\downarrow} c\pdg_{i,{\sf m},\uparrow}) \nonumber \\
S^{\sf y}_{i} &= -\frac{i}{2} \sum_{m={\sf a,b}} (c\dg_{i,{\sf m},\uparrow} c\pdg_{i,{\sf m},\downarrow}-c\dg_{i,{\sf m},\downarrow} c\pdg_{i,{\sf m},\uparrow}) \nonumber \\
S^{\sf z}_{i} &= \frac{1}{2} \sum_{m={\sf a,b}} (c\dg_{i,{\sf m},\uparrow} c\pdg_{i,{\sf m},\uparrow}-c\dg_{i,{\sf m},\downarrow} c\pdg_{i,{\sf m},\downarrow}),
\end{align}
and,
\begin{align}
T^{\sf x}_{i} &= \frac{1}{2} \sum_{\sigma=\uparrow,\downarrow} (c\dg_{i,{\sf a},\sigma} c\pdg_{i,{\sf b},\sigma}+c\dg_{i,{\sf b},\sigma} c\pdg_{i,{\sf a},\sigma}) \nonumber \\
T^{\sf y}_{i} &= -\frac{i}{2} \sum_{\sigma=\uparrow,\downarrow} (c\dg_{i,{\sf a},\sigma} c\pdg_{i,{\sf b},\sigma}-c\dg_{i,{\sf b},\sigma} c\pdg_{i,{\sf a},\sigma}) \nonumber \\
T^{\sf z}_{i} &= \frac{1}{2} \sum_{\sigma=\uparrow,\downarrow} (c\dg_{i,{\sf a},\sigma} c\pdg_{i,{\sf a},\sigma}-c\dg_{i,{\sf b},\sigma} c\pdg_{i,{\sf b},\sigma}).
\end{align}

Performing second order perturbation theory results in,
\begin{align}
&\mathcal{H}_{\sf ST} = \frac{4(t^\prime)^2}{U} \sum_{\langle ij \rangle} 
\left\{
-\frac{1}{1+J/U} 
\mathcal{P}_{ij}^{S=0}
\left[ \frac{2t}{t^\prime} {\bf T}_i\cdot {\bf T}_j  
-\frac{4t}{t^\prime} T^{\sf y}_i T^{\sf y}_j \right. \right.\nonumber \\
& \quad +\left( 1- t/t^\prime \right)^2 ({\bf n}_{ij}\cdot {\bf T}_i)({\bf n}_{ij} \cdot {\bf T}_j) \nonumber \\
& \quad \left. -\frac{1}{2} \left( 1- \left( t/t^\prime \right)^2 \right) ({\bf n}_{ij}\cdot {\bf T}_i + {\bf n}_{ij} \cdot {\bf T}_j)
+\frac{1}{4} \left( 1+ \left(t/t^\prime\right)^2 \right)
\right] \nonumber \\
&-\frac{1}{1-J/U} 
\mathcal{P}_{ij}^{S=0}
\left[ \frac{4t}{t^\prime} T^{\sf y}_i T^{\sf y}_j \right. \nonumber \\
& \quad \left.-\frac{1}{2} \left( 1- \left(t/t^\prime\right)^2 \right) ({\bf n}_{ij}\cdot {\bf T}_i + {\bf n}_{ij} \cdot {\bf T}_j)
+\frac{1}{2} \left( 1+ \left(t/t^\prime\right)^2 \right)
\right] \nonumber \\
& +\frac{1}{1-3J/U} 
\mathcal{P}_{ij}^{S=1}
\left[ \frac{2t}{t^\prime} {\bf T}_i\cdot {\bf T}_j  \right. \nonumber \\
&\quad \left. \left. +\left( 1- t/t^\prime \right)^2 ({\bf n}_{ij}\cdot {\bf T}_i)({\bf n}_{ij} \cdot {\bf T}_j) 
-\frac{1}{4} \left( 1+ \left(t/t^\prime\right)^2 \right)
\right]
\right\},
\label{eq:HST-honeycomb}
\end{align}
where, 
\begin{align}
\mathcal{P}_{ij}^{S=0} = \frac{1}{4} -{\bf S}_i\cdot {\bf S}_j , \quad
\mathcal{P}_{ij}^{S=1} = \frac{3}{4} + {\bf S}_i\cdot {\bf S}_j ,
\end{align}
are the spin singlet and triplet projection operators, and the reparametrisation,
\begin{align}
J_{\sf H} = 2J_{\sf p} = J, \qquad
U = 2\tilde{U} +J,
\label{eq:JU-reparam}
\end{align}
has been made.
The vectors ${\bf n}_{ij}$ are different for {\sf A}, {\sf B} and {\sf C} bonds (see Fig.~\ref{fig:mapping} for bond labelling) and given by,
\begin{align}
{\bf n}_{ij\in {\sf A}} &= (0,0,1) \nonumber \\
{\bf n}_{ij\in {\sf B}} &= \left( \frac{\sqrt{3}}{2},0,-\frac{1}{2} \right) \nonumber \\
{\bf n}_{ij\in {\sf C}} &=  \left( -\frac{\sqrt{3}}{2},0,-\frac{1}{2} \right).
\label{eq:nvectors}
\end{align}

It is interesting to compare $\mathcal{H}_{\sf ST}$ [Eq.~(\ref{eq:HST-honeycomb})] to the superexchange Hamiltonian derived in Eq.~(1) of Ref.~[\onlinecite{nasu13}].
Qualitatively, the two Hamiltonians contain the same combinations of spin and orbital operators -- those allowed by the symmetry of the honeycomb lattice.
However, the coefficients in front of these terms are parametrised differently. 
One source of difference is the form of the pair hopping term.
In $\mathcal{H}^{\sf coul}$ [Eq.~\ref{eq:Hcoul}] the $J_{\sf p}$ term is crucial for splitting the multiplet of doubly occupied Cu states, but is omitted in Ref.~[\onlinecite{nasu13}].
Another source of difference is the inclusion in Ref.~[\onlinecite{nasu13}] of ``d-p-d'' hopping, which describes superexchange interactions in which the O p-orbitals rather than the Cu d-orbitals are doubly occupied in the intermediate state.
Here we ignore this type of hopping with respect to the ``d-d'' hopping in which the Cu d-orbitals are doubly occupied.

The Hamiltonian $\mathcal{H}_{\sf ST}$ [Eq.~(\ref{eq:HST-honeycomb})] is {\sf SU(2)} symmetric in the spin degree of freedom, but highly anisotropic in terms of the orbital degree of freedom.
Below we use numerics to determine the ground state of $\mathcal{H}_{\sf ST}$ on small clusters, varying the two free parameters $t/t^\prime$ and $J/U$ to generate a phase diagram.
However, before describing the numerical results, it is first useful to study $\mathcal{H}_{\sf ST}$ analytically.
We demonstate a mapping connecting $t/t^\prime=1$ to $t/t^\prime=-1$, and then consider some highly frustrated points and lines with enhanced symmetry, which are particularly important for the ground state phase diagram.


\subsection{Mapping t/t$^\prime$=1 $\to$ t/t$^\prime$=-1}
\label{sec:mapping}


A canonical transformation relates $\mathcal{H}_{\sf ST}$ [Eq.~(\ref{eq:HST-honeycomb})] at $t/t^\prime=1$ and $t/t^\prime=-1$.
As a consequence, the energy eigenspectrum is invariant under the transformation $t/t^\prime=1 \to t/t^\prime=-1$.
The mapping involves an orbital rotation with 6-sublattice structure, and is given in Table~\ref{tab:mapping}.

\begin{table}
\begin{center}
\footnotesize
  \begin{tabular}{| c ||  c | c | c |  }
    \hline 
\multirow{2}{*}{sublattice $i$}  & 
\multirow{2}{*}{$T_i^{\sf x,in}$} & 
\multirow{2}{*}{$T_i^{\sf y,in}$}  &
\multirow{2}{*}{$T_i^{\sf z,in}$}   \\ 
&  &  &   \\ \hline \hline
\multirow{2}{*}{1}  & 
\multirow{2}{*}{$T_1^{\sf x}$} & 
\multirow{2}{*}{$T_1^{\sf y}$}  &
\multirow{2}{*}{$T_1^{\sf z}$}   \\ 
&  &  &   \\ \hline
\multirow{2}{*}{2}  & 
\multirow{2}{*}{$-T_2^{\sf x}$} & 
\multirow{2}{*}{$-T_2^{\sf y}$}  &
\multirow{2}{*}{$T_2^{\sf z}$}    \\ 
&  &  &  \\ \hline
\multirow{2}{*}{3}  & 
\multirow{2}{*}{$-\frac{1}{2}T_3^{\sf x} \pm \frac{\sqrt{3}}{2}T_3^{\sf z}$} & 
\multirow{2}{*}{$T_3^{\sf y}$}  &
\multirow{2}{*}{$\mp \frac{\sqrt{3}}{2}T_3^{\sf x}-\frac{1}{2}T_3^{\sf z}$}    \\ 
&  &  &  \\ \hline
\multirow{2}{*}{4}  & 
\multirow{2}{*}{$\frac{1}{2}T_4^{\sf x}-\frac{\sqrt{3}}{2}T_4^{\sf z}$} & 
\multirow{2}{*}{$-T_4^{\sf y}$}  &
\multirow{2}{*}{$-\frac{\sqrt{3}}{2}T_4^{\sf x}-\frac{1}{2}T_4^{\sf z}$}    \\ 
&  &  &  \\ \hline
\multirow{2}{*}{5}  & 
\multirow{2}{*}{$-\frac{1}{2}T_5^{\sf x} \mp \frac{\sqrt{3}}{2}T_5^{\sf z}$} & 
\multirow{2}{*}{$T_5^{\sf y}$}  &
\multirow{2}{*}{$\pm \frac{\sqrt{3}}{2}T_5^{\sf x}-\frac{1}{2}T_5^{\sf z}$}   \\ 
&  &  &  \\ \hline
\multirow{2}{*}{6}  & 
\multirow{2}{*}{$\frac{1}{2}T_6^{\sf x}+\frac{\sqrt{3}}{2}T_6^{\sf z}$} & 
\multirow{2}{*}{$-T_6^{\sf y}$}  &
\multirow{2}{*}{$\frac{\sqrt{3}}{2}T_6^{\sf x}-\frac{1}{2}T_6^{\sf z}$}    \\ 
&  &  & \\ \hline
    \end{tabular}
\end{center} 
\caption{\footnotesize{
Orbital transformations used in the mapping \mbox{$t/t^\prime=1 \to t/t^\prime=-1$} (upper sign) and \mbox{$t/t^\prime \to -t/t^\prime$} (lower sign).
The \mbox{$t/t^\prime \to -t/t^\prime$} mapping is defined for the 6 and 18 site clusters with periodic boundary conditions [see also Fig.~\ref{fig:mapping}].
The lattice is divided into 6 sublattices such that each hexagon contains one site in each sublattice.
The orbital pseudospin operator on sublattice $i$ is mapped from an initial value $T_i^{\alpha,{\sf in}} \to a T_i^{\sf x} + bT_i^{\sf y} + cT_i^{\sf z}$, where $a$, $b$ and $c$ are a set of coefficients obeying $\sqrt{a^2+b^2+c^2}=1$.
}}
\label{tab:mapping}
\end{table}

For the 6 and 18 site clusters with periodic boundary conditions considered below, it is possible to make the more general mapping $t/t^\prime \to -t/t^\prime$ [see Fig.~\ref{fig:mapping}].
This provides a strong constraint on the symmetry of the ground state phase diagram, and allows the nature of phases with $t/t^\prime<0$ to be deduced from the corresponding $t/t^\prime>0$ phase or vice versa.
The transformation involves an orbital rotation with 6-sublattice structure and a compensatory restructuring of the lattice. 
The orbital rotation is shown in Table~\ref{tab:mapping} and the lattice restructuring in Fig.~\ref{fig:mapping}.
The lattice restructuring relies on mapping orbits of the cluster onto hexagons, and therefore does not generalise to larger sizes.


\subsection{Special points and lines}
\label{sec:SU4andcompass}


The Hamiltonian $\mathcal{H}_{\sf ST}$ [Eq.~(\ref{eq:HST-honeycomb})] contains a special point with {\sf SU(4)} symmetry, and also a highly frustrated line on which the orbital part of $\mathcal{H}_{\sf ST}$ reduces to the compass model.
These are very important for the ground state phase diagram.


\subsubsection{SU(4) point}


$\mathcal{H}_{\sf ST}$ [Eq.~(\ref{eq:HST-honeycomb})] is {\sf SU(4)} symmetric for the parameters, 
\begin{align}
t=t^\prime, \quad 
J=0.
\end{align}
At this point the Hamiltonian is given by, 
\begin{align}
\mathcal{H}_{\sf ST}^{\sf SU(4)} &= \frac{16t^2}{U} \sum_{\langle ij \rangle} 
\left(  {\bf T}_i\cdot {\bf T}_j +\frac{1}{4} \right)
\left(  {\bf S}_i\cdot {\bf S}_j +\frac{1}{4} \right),
\label{eq:HSU4}
\end{align}
and one can freely rotate between spin and orbital degrees of freedom.
Detailed studies of $\mathcal{H}_{\sf ST}^{\sf SU(4)}$ have been carried out in Ref.~[\onlinecite{corboz12}], and it was shown that the ground state is a spin-orbital liquid.
There is good evidence that this spin-orbital liquid has algebraically decaying correlations.

Below, we find that the ground state phase diagram of $\mathcal{H}_{\sf ST}$ [Eq.~(\ref{eq:HST-honeycomb})] has a sizeable region that continuously connects to the {\sf SU(4)} point.
We therefore interpret this phase as a spin-orbital liquid phase.
The point $t=-t^\prime$, $J=0$ is connected to the {\sf SU(4)} point by the $t/t^\prime=1 \to t/t^\prime=-1$ mapping described in Section~\ref{sec:mapping}, and thus has a hidden {\sf SU(4)} symmetry.

\begin{figure*}[ht]
\centering
\includegraphics[width=0.85\textwidth]{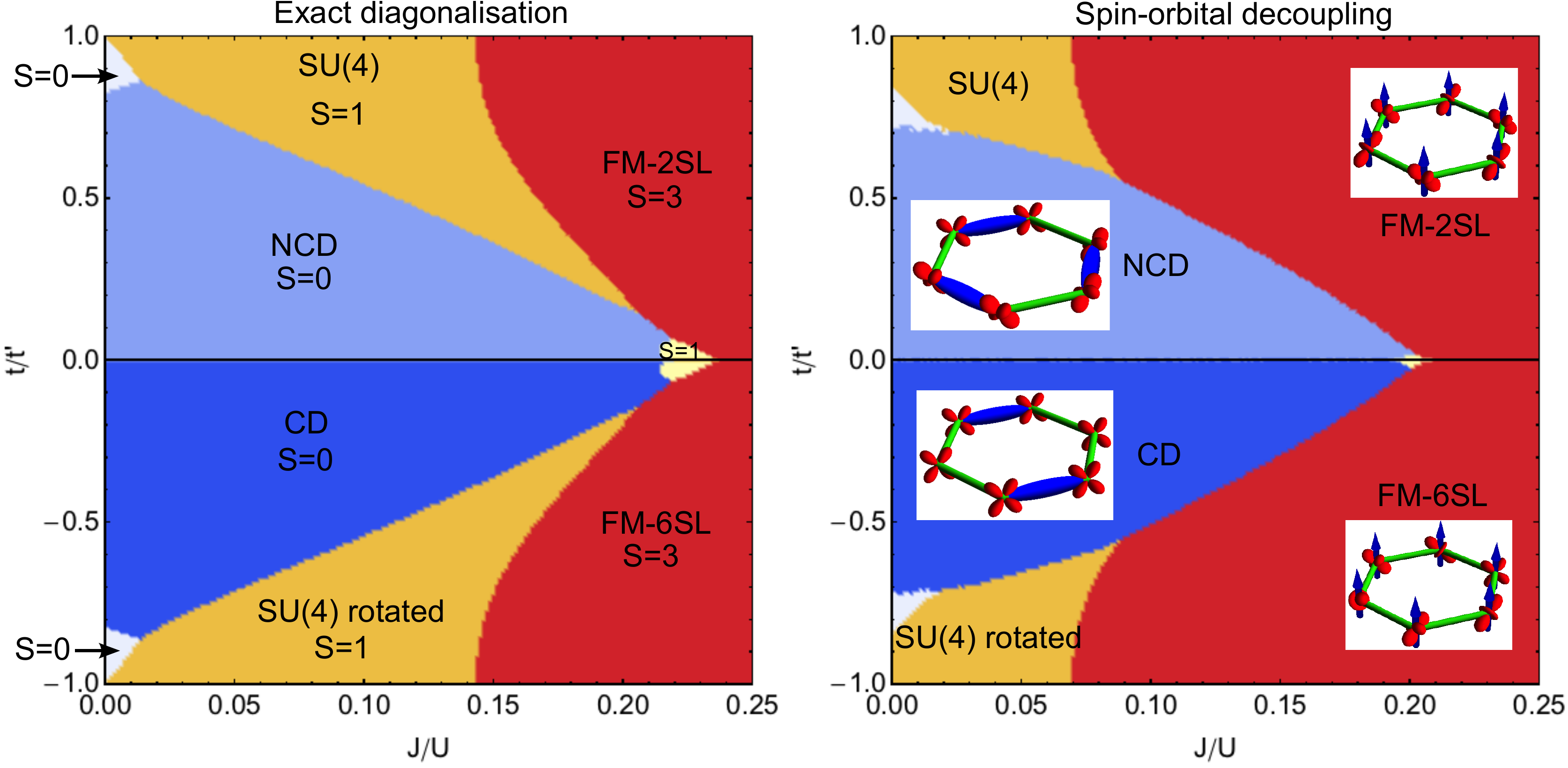}
\caption{\footnotesize{
Phase diagram for $\mathcal{H}_{\sf ST}$ [Eq.~\ref{eq:HST-honeycomb}] on the honeycomb lattice, as calculated from (a) exact diagonalisation and (b) spin-orbital decoupling on a 6-site cluster.
Illustrations of phases show spins in blue, with ellipsoids on bonds representing spin-singlets, and orbitals in red.
At $t/t^\prime=1$ and $J/U=0$, $\mathcal{H}_{\sf ST}$ is {\sf SU(4)} symmetric, and the two regions continuously connected to this point  are labelled as {\sf SU(4)} phases.
The gold region has $S_{\sf tot}=1$, $T_{\sf tot}\approx 0$ and the white region $S_{\sf tot}=0$, $T_{\sf tot}\approx 1$.
The light blue region has $S_{\sf tot}=0$, non-collinear ordering of nearest-neighbour spin dimers (NCD) and 3-sublattice orbital order of $d^{\sf x^2-y^2}$ type.
The darker blue region has $S_{\sf tot}=0$, collinear ordering of nearest-neighbour spin dimers (CD) and ferro-orbital order of $d^{\sf x^2-y^2}$ type.
The red region is ferromagnetic, with antiferro-orbital ordering for $t/t^\prime>0$ and 6-sublattice orbital order for $t/t^\prime<0$.
The orbitals alternate between $d^{\sf x^2-y^2}$ and $d^{\sf 3z^2-r^2}$ type on neighbouring sites. 
Finally, the small yellow region close to $t/t^\prime=0$ is an intermediary spin state with $S_{\sf tot}=1$. 
}}
\label{fig:honeycomb-n6-pd}
\end{figure*}
%


\subsubsection{The line t/t$^\prime$=0}


We now turn to the line $t/t^\prime=0$, which is highly frustrated.
The ground state is disordered, and can be understood in terms of the orbital compass model.
At finite $t/t^\prime$, small ${\bf T}_i\cdot {\bf T}_j$ perturbations break this degeneracy, and select an orbitally ordered state.

The simplest illustration occurs at large $J/U$, where the spins align ferromagnetically. 
Setting $t/t^\prime=0$ and ${\bf S}_i\cdot {\bf S}_j=1/4$ on every bond, leads to the Hamiltonian,
\begin{align}
\mathcal{H}^{\sf comp}_{\sf T} &= \frac{4(t^\prime)^2}{U} \frac{1}{1-3J/U} 
 \sum_{\langle ij \rangle} 
\left[ ({\bf n}_{ij}\cdot {\bf T}_i)({\bf n}_{ij} \cdot {\bf T}_j) -\frac{1}{4}  \right].
\label{eq:orbcompass}
\end{align}
This model was studied in Ref.~[\onlinecite{nasu08}].
Classically, the ground state degeneracy of 2-dimensional compass models scales as $\mathcal{O}(\sqrt{N})$, where $N$ is the number of lattice sites\cite{dorier05}.
The classical ground states consist of all possible dimer coverings of the honeycomb lattice, where a dimer corresponds to a minimum energy nearest-neighbour bond.
For $J/U<1/3$, a minimum energy bond requires one orbital pseudospin to align parallel to ${\bf n}_{ij}$ and the other antiparallel.
Only 1/3 of bonds can minimise their energy, and the remaining bonds are frustrated.
The quantum ground state is well described by a linear superposition of the classical ground states\cite{nasu08}.

At small $t/t^\prime$ the Hamiltonian also includes a Heisenberg orbital interaction, and is given by,
\begin{align}
\mathcal{H}_{\sf T} &= \frac{4(t^\prime)^2}{U} \frac{1}{1-3J/U} 
 \sum_{\langle ij \rangle} 
\left[ \frac{2t}{t^\prime} {\bf T}_i\cdot {\bf T}_j  \right. \nonumber \\
 & \left.+\left( 1- t/t^\prime \right)^2 ({\bf n}_{ij}\cdot {\bf T}_i)({\bf n}_{ij} \cdot {\bf T}_j) 
-\frac{1}{4} \left( 1+ \left(t/t^\prime \right)^2 \right)
\right].
\label{eq:HTfm}
\end{align}
At the classical level, an infinitesimal Heisenberg term breaks the degeneracy of $\mathcal{H}^{\sf comp}_{\sf T}$ [Eq.~(\ref{eq:orbcompass})], resulting in an orbitally ordered ground state.
The ground state chosen is the one giving the best energy on the 2/3 of bonds that frustrate the compass term.
The detailed nature of the ground state depends on the sign of $t/t^\prime$ and whether $J/U<1/3$ or $J/U>1/3$.
For the quantum Hamiltonian, exact diagonalisation of an 18-site cluster shows orbital order at $t/t^\prime=10^{-3}$, the lowest value checked.
This orbital order is consistent with that expected from classical considerations.

A qualitatively similar analysis can be made at small $J/U$, where the spins no longer order ferromagnetically.
At $t/t^\prime=0$, minimum energy bonds correspond to aligning both orbital pseudospins anti-parallel to ${\bf n}_{ij}$ and forming a spin singlet.  
The preference for placing orbital pseudospins antiparallel to ${\bf n}_{ij}$, as opposed to parallel, is due to the sign in front of the \mbox{${\bf n}_{ij}\cdot {\bf T}_i + {\bf n}_{ij} \cdot {\bf T}_j$} term, which acts like an orbital magnetic field.
A low-energy variational subspace is formed by covering 1/3 of the bonds of the honeycomb lattice with minimum energy bonds.
At small $t/t^\prime$ the orbital Heisenberg perturbation selects an orbitally ordered ground state, which in turn leads to a valence bond solid ground state in the spin-sector.
The ground states at $t/t^\prime<0$ and $t/t^\prime>0$ are different, but related by the transformation described in Section~\ref{sec:mapping} above.


\subsection{Ground state phase diagram for the 6-site cluster}


We now numerically study the ground state phase diagram of $\mathcal{H}_{\sf ST}$ [Eq.~\ref{eq:HST-honeycomb}] on small clusters as a function of $t/t^\prime$ and $J/U$.
For the 6-site cluster with periodic boundary conditions, $\mathcal{H}_{\sf ST}$ already shows all the important features found on larger clusters, and can be fully diagonalised.

We study the 6-site cluster both by exact diagonalisation and also using a mean field approximation based on decoupling spin and orbital degrees of freedom.
 This method was pioneered in Refs.~[\onlinecite{mila98,vernay04}] and is explained in detail in Appendix~\ref{App:SOdecoup}.
It involves factorising the wavefunction into spin and orbital components, and diagonalising the two components self consistently. 
The motivation for using this method on the 6-site cluster is twofold.
Firstly we show that it compares well to exact diagonalisation, suggesting that it is trustworthy on larger clusters for which exact diagonalisation is computationally expensive.
Secondly it provides a simple way of identifying the nature of the phases.

The ground state phase diagram on the 6-site cluster is shown in Fig.~\ref{fig:honeycomb-n6-pd}.
Exactly at the {\sf SU(4)} point there is a 24-fold degeneracy of the ground state.
This consists of 12 states with $S_{\sf tot}=0$ and $T_{\sf tot}=1$ and 12 with $S_{\sf tot}=1$ and $T_{\sf tot}=0$.
Away from the {\sf SU(4)} point $T_{\sf tot}$ is no longer a good quantum number, but for small deviations it can still be used to classify the phases.
The white phase in Fig.~\ref{fig:honeycomb-n6-pd} has $S_{\sf tot}=0$ and $T_{\sf tot}\approx 1$, while the gold phase has $S_{\sf tot}=1$ and $T_{\sf tot}\approx 0$.
These phases have a 2-fold degenerate ground state, except for the line $t/t^\prime=1$, where there is a 4-fold degeneracy.
The phases connecting to $t/t^\prime=-1$ and $J/U=0$ are related by the mapping given in Section~\ref{sec:mapping}.

The blue phases that dominate the centre of the phase diagram are orbitally ordered, and the spins form a valence bond solid of nearest-neighbour spin singlets, consistent with the analytic arguments put forward in Section~\ref{sec:SU4andcompass}.
For $t/t^\prime<0$ there is $d^{\sf x^2-y^2}$ type ferro-orbital order and a collinear arrangement of nearest-neighbour singlet bonds (CD state).
For $t/t^\prime>0$ there is antiferro-orbital order involving $d^{\sf x^2-y^2}$, $d^{\sf y^2-z^2}$ and $d^{\sf z^2-x^2}$ orbitals (henceforth denoted as $d^{\sf x^2-y^2}$-type orbitals) and a non-collinear ``Kekul{\' e}'' arrangement of nearest-neighbour singlet bonds (NCD state).

At large $J/U$ the Hund's rule coupling favours ferromagnetic order.
This coexists with orbital order, which involves alternating orbitals of $d^{\sf x^2-y^2}$ type and $d^{\sf 3z^2-r^2}$ type.
For $t/t^\prime<0$ this orbital order has 6 sublattice structure, while for For $t/t^\prime>0$ it has 2 sublattices.

Finally there is an intermediate spin phase with $S=1$ at $t/t^\prime\approx 0$.
This involves one spin triplet bond and two spin singlet bonds.

It is evident from Fig.~\ref{fig:honeycomb-n6-pd} that the phase diagram calculated via the spin-orbital decoupling mean-field approximation compares well to that calculated by exact diagonalisation.
This is especially true if one is interested in small $J/U$, as is the case for Ba$_3$CuSb$_2$O$_9$.
The largest discrepancy occurs in the regions surrounding $J/U=0.15$, $t/t^\prime=\pm 0.5$.
These regions are assigned to the {\sf SU(4)} phase in the exact diagonalisation phase diagram but to the ferromagnetic phase in the spin-orbital decoupling phase diagram.
The similarity of the two phase diagrams indicates that it is reasonable to use the mean field approximation on larger clusters, where full diagonalisation is prohibitively expensive.


\subsection{Ground state phase diagram for the 18-site cluster}


%
\begin{figure}[h]
\centering
\includegraphics[width=0.49\textwidth]{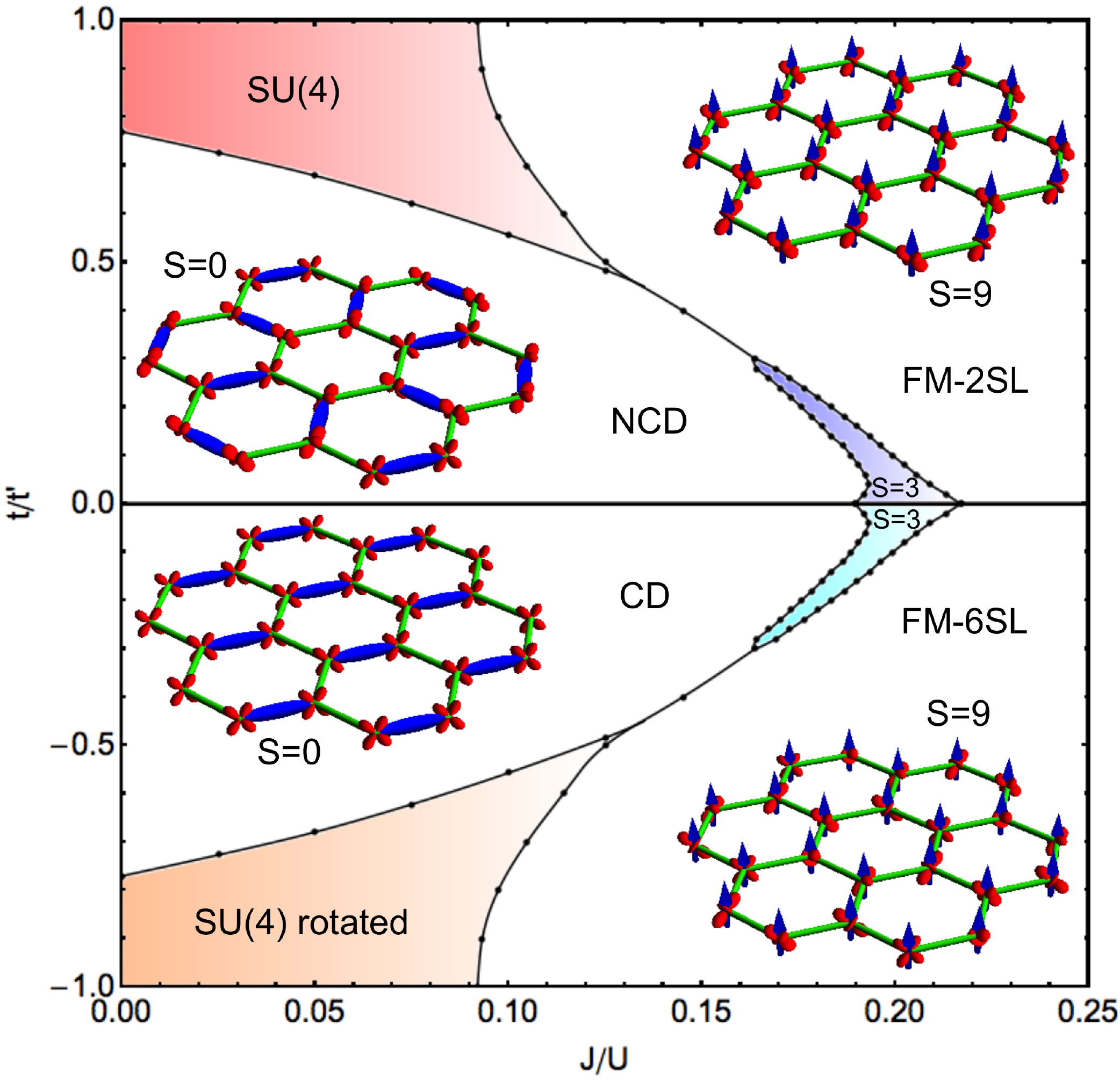}
\caption{\footnotesize{
Phase diagram for $\mathcal{H}_{\sf ST}$ [Eq.~\ref{eq:HST-honeycomb}] on the honeycomb lattice, as calculated within the spin-orbital decoupling scheme on an 18-site cluster.
Illustrations of phases show spins in blue, with ellipsoids on bonds representing spin-singlets, and orbitals in red.
The {\sf SU(4)} phase contains the {\sf SU(4)} point at $t=t^\prime$ and $J/U=0$.
The rotated {\sf SU(4)} phase is related by the orbital mapping described in Section~\ref{sec:mapping}.
The non-collinear dimer phase (NCD) is an $S_{\sf tot}=0$ phase with nearest-neighbour spin singlets crystallised in a non-collinear pattern and $d^{\sf x^2-y^2}$ type orbital order.
The collinear dimer (CD) phase involves crystallisation of spin dimers in a collinear pattern, with $d^{\sf x^2-y^2}$ type ferro-orbital order.
At large $J/U$ the spin configuration is ferromagnetic.
For $t/t^\prime>0$ there is antiferro-orbital order with an alternation of $d^{\sf x^2-y^2}$ and $d^{\sf 3z^2-r^2}$ type orbitals, while for $t/t^\prime<0$ the orbital order has a 6-sublattice structure.
Finally there is an intermediate spin region with $S_{\sf tot}=3$.
}}
\label{fig:honeycomb-n18-pd}
\end{figure}

For the 18-site cluster with periodic boundary conditions we use the spin-orbital decoupling method [see Appendix~\ref{App:SOdecoup}] to map out the ground state phase diagram.
This is shown in Fig.~\ref{fig:honeycomb-n18-pd}, and there are no qualitative changes from the 6-site cluster.
Compared to the spin-orbital decoupling method for the 6-site cluster, the {\sf SU(4)} phase survives to higher $J/U$, at the expense of the ferromagnetic phase.
Deep inside the {\sf SU(4)} region the mean field approximation is not so reliable, and we do not attempt to split this phase into different spin sectors.
The intermediate $S_{\sf tot}=3$ phase involves a ferromagnetic chain surrounded by spin singlets.
It is likely that this phase does not survive in the thermodynamic limit.


\section{Ground state of spin-orbital model on the decorated honeycomb lattice}
\label{sec:dechoneycomb}


%
\begin{figure}[ht]
\centering
\includegraphics[width=0.3\textwidth]{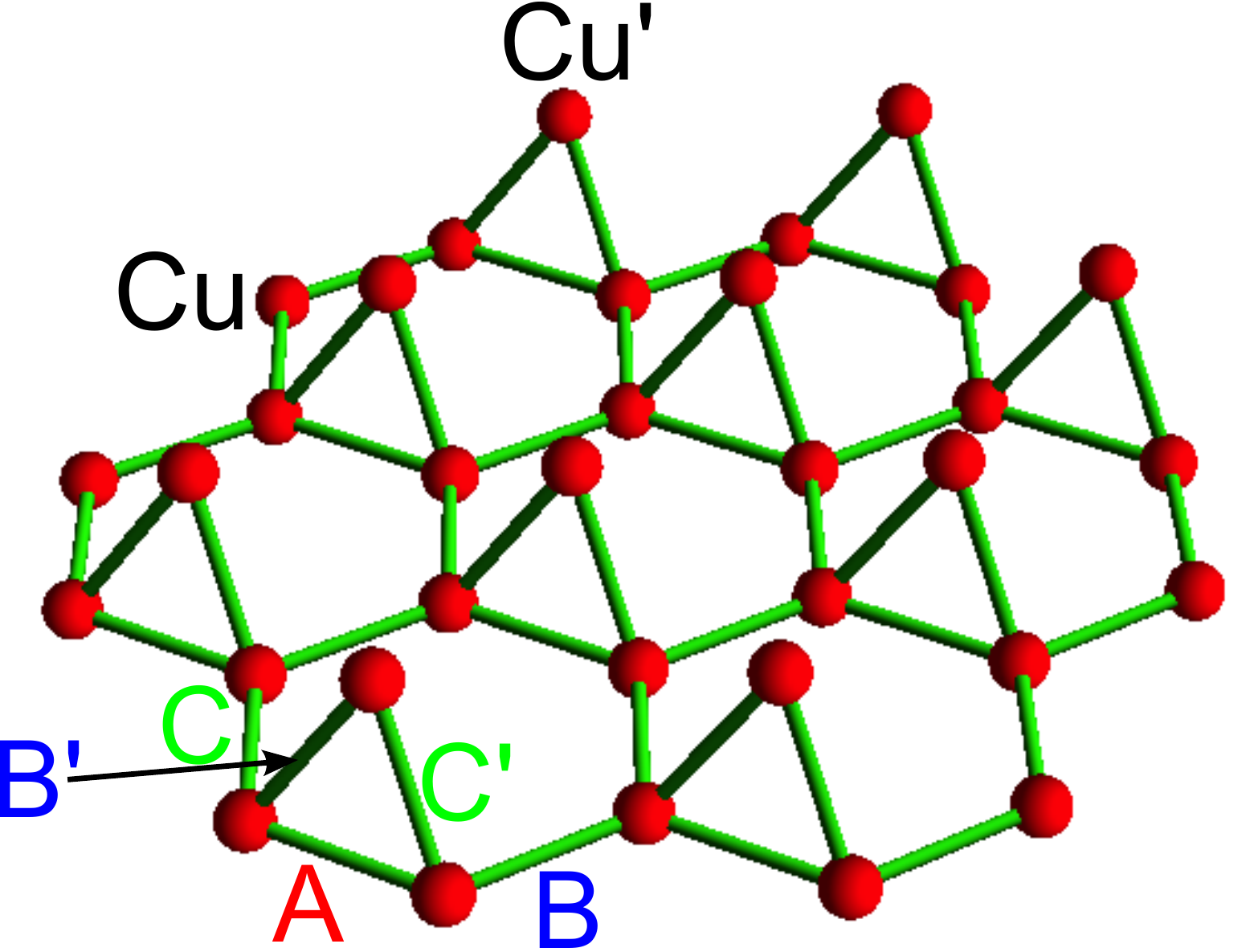}
\caption{\footnotesize{
The decorated honeycomb lattice.
A honeycomb lattice of Cu ions is decorated by the addition of Cu$^\prime$ ions, forming \mbox{Cu-Cu$^\prime$-Cu} isoceles triangles.
Bonds on the honeycomb lattice are labelled {\sf A}, {\sf B} and {\sf C}, as in Fig.~\ref{fig:mapping}, while Cu-Cu$^\prime$ bonds are labelled ${\sf B}^\prime$ and ${\sf C}^\prime$.
}}
\label{fig:dechoney}
\end{figure}
\begin{figure*}[ht]
\centering
\includegraphics[width=0.98\textwidth]{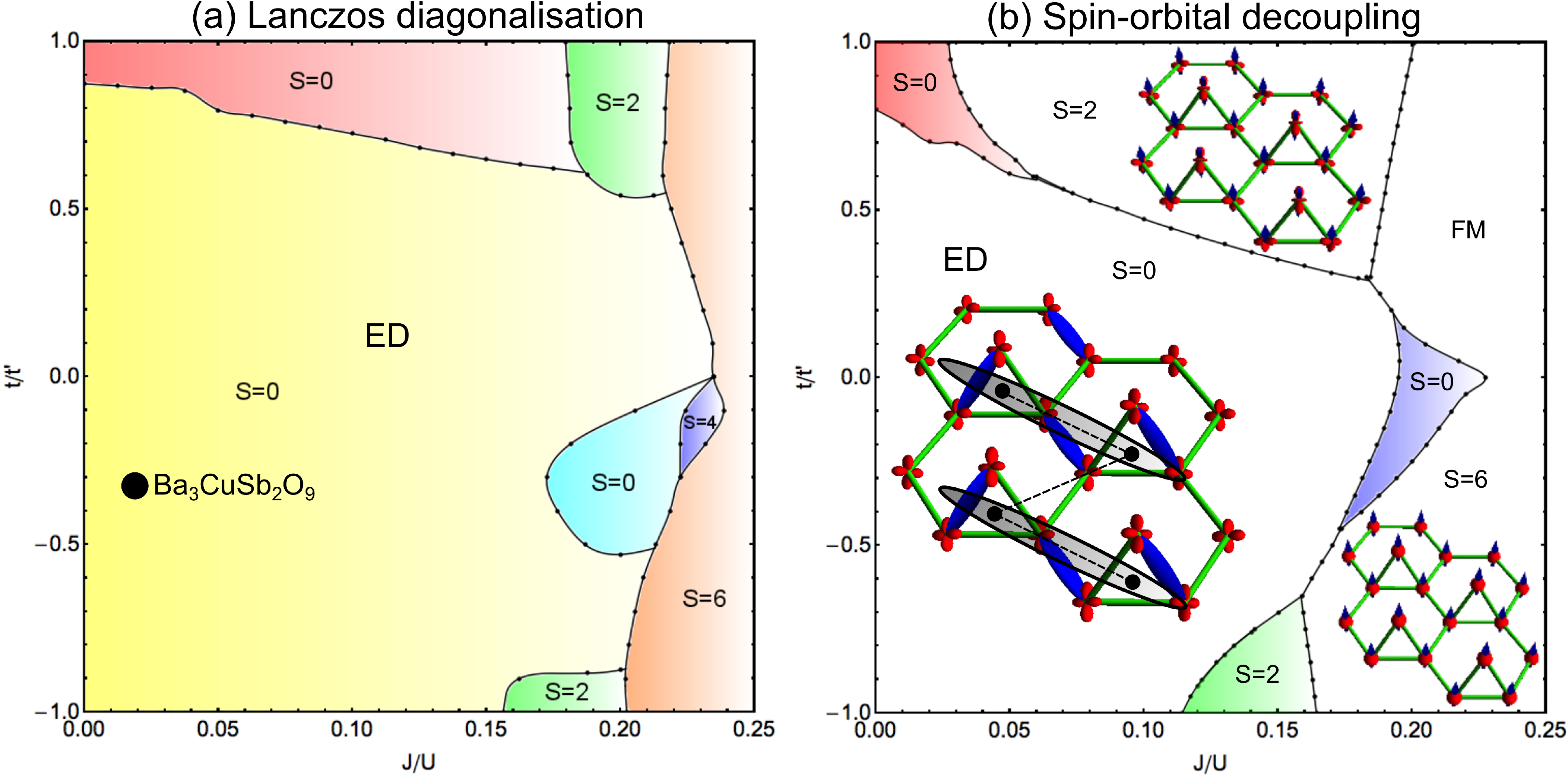}
\caption{\footnotesize{
Phase diagram for \mbox{$\mathcal{H}_{\sf ST} + \mathcal{H}_{\sf ST}^\prime$} [Eqs.~(\ref{eq:HST-honeycomb},\ref{eq:H-CuCuprime})] on the decorated honeycomb lattice, calculated using (a) Lanczos diagonalisation and (b) spin-orbital decoupling on a 12 site cluster.
Parameters $t_{\sf a}/t^\prime=2/3$, $t_{\sf b}/t^\prime=0$ and $t_{\sf ab}/t^\prime=-1/\sqrt{3}$ are used.
At small $J/U$ the phase diagram is dominated by an $S=0$ phase (yellow, ED) in which 3 nearest neighbour spin singlets (blue ellipses) form on 6-site clusters.
This cluster of 3 singlet bonds and associated orbital order (shown in red) can be thought of as a dimer (black ellipse) on an emergent square lattice (see Fig.~\ref{fig:6sitedimer} and Fig.~\ref{fig:effsqlat}), and we label it the emergent dimer phase (ED).
The 12-site cluster is too small to determine the ordering pattern of these dimers in the thermodynamic limit.
The red $S=0$ phase has the same spin-singlet pattern, but a different orbital state.
Other phases include a pair of $S=2$ phases (green), which consist of ferromagnetic spins on the honeycomb lattice aligned antiparallel to the spins on the Cu$^\prime$ sites.
Orbital order on the honeycomb lattice is of $d^{\sf x^2-y^2}$ type, while orbitals on the Cu$^\prime$ sites are $d^{\sf 3z^2-r^2}$ type.
For large $J/U$ the spins order ferromagnetically (orange, FM), and the orbital ordering is likely incommensurate.
There are also intermediate phases (light and dark blue), which do not match between the exact diagonalisation and the spin-orbital decoupling. 
The $S=0$ phase (dark blue) in the spin-orbital decoupling phase diagram involves  ferromagnetically aligned chains that order antiferromagnetically with neighbouring chains.
This has a significant overlap with the first excited state of the $S=4$ (dark blue) phase calculated via exact diagonalisation.
}}
\label{fig:dechoneycomb-n12-pd}
\end{figure*}

We now turn to the case of the decorated honeycomb lattice.
It is expected in Ba$_3$CuSb$_2$O$_9$ that there are two dominant Cu-O-O-Cu superexchange pathways with comparable hopping amplitudes\cite{nakatsuji12}: one between Cu ions in the same plane (path 1 in Fig.~\ref{fig:Ba3CuSb2O9-structure}) and one between Cu ions in neighbouring bilayers (path 2 in Fig.~\ref{fig:Ba3CuSb2O9-structure}).
In consequence, one is lead to consider a honeycomb lattice of Cu ions, decorated by out of plane Cu$^\prime$ ions, as shown in Fig.~\ref{fig:dechoney}.
The Cu-Cu bond length is measured as 5.81\AA, while the Cu-Cu$^\prime$ bond length is 5.61\AA \cite{nakatsuji12}.
The interesting question is whether the addition of the extra Cu$^\prime$ ions significantly changes the ground state phase diagrams shown in Fig.~\ref{fig:honeycomb-n6-pd} and Fig.~\ref{fig:honeycomb-n18-pd}.


\subsection{Microscopic model}


It is first necessary to consider how the addition of the Cu$^\prime$ sites changes the microscopic model, $\mathcal{H}^{\sf Hub}$ [Eq.~(\ref{eq:Hhub})].
On the decorated honeycomb lattice we write,
\begin{align}
\mathcal{H}^{\sf Hub} = \mathcal{H}^{\sf hop} + \mathcal{H}^{\prime {\sf hop}} + \mathcal{H}^{\sf coul} ,
\label{eq:Hhubdec}
\end{align}
where $\mathcal{H}^{\prime {\sf hop}}$ describes Cu-Cu$^\prime$  hopping.
We make the assumption that the hopping parameters within the honeycomb lattice, described by $\mathcal{H}^{\sf hop}$ [Eq.~(\ref{eq:Hhop})], are unchanged.
This is an approximation, since the Cu$^\prime$ sites break the $C_3$ symmetry of the honeycomb lattice, and therefore invalidate the relationship between hopping on {\sf A}, {\sf B} and {\sf C} bonds used in Section~\ref{sec:honeycomb} [see Eq.~(\ref{eq:crotation})].
However, we expect deviation from the $C_3$ symmetric case to be small.
We also consider the same Coulomb Hamiltonian, $\mathcal{H}^{\sf coul}$ [Eq.~(\ref{eq:Hcoul})] for both Cu and Cu$^\prime$ sites.

The hopping Hamiltonian on ${\sf B}^\prime$ bonds (see Fig.~\ref{fig:dechoney} for bond labelling) can be written as, 
\begin{align}
\mathcal{H}_{\sf B^\prime}^{\prime{\sf hop}} =& 
-t_{\sf a} \sum_{\sigma} c\dg_{i,{\sf a},\sigma} c\pdg_{j,{\sf a},\sigma}
-t_{\sf b} \sum_{\sigma} c\dg_{i,{\sf b},\sigma} c\pdg_{j,{\sf b},\sigma} \nonumber \\
&-t_{\sf ab} \sum_{\sigma} c\dg_{i,{\sf b},\sigma} c\pdg_{j,{\sf a},\sigma}
-t_{\sf ab} \sum_{\sigma} c\dg_{i,{\sf a},\sigma} c\pdg_{j,{\sf b},\sigma}
+\mathrm{H.c.}
\label{eq:Hhop-Cuprime}
\end{align}
and $\mathcal{H}_{\sf C^\prime}^{\prime{\sf hop}}$ follows from the mirror symmetry transformation, Eq.~(\ref{eq:mirsymtransformation}).


\subsection{Effective spin-orbital Hamiltonian}


An effective spin-orbital model can be derived from the microscopic Hamiltonian $\mathcal{H}^{\sf Hub}$ [Eq.~(\ref{eq:Hhubdec})] using second order perturbation theory, as in Section~\ref{sec:honeycomb}.
We are lead to consider $\mathcal{H}_{\sf ST} + \mathcal{H}_{\sf ST}^\prime$, where,
\begin{align}
&\mathcal{H}_{\sf ST}^\prime= 
-\frac{4}{U}{\sum_{ij}}^\prime
\left\{
\frac{1}{1+J/U}
 \mathcal{P}_{ij}^{\sf S=0}
 \left[
2(t_{\sf a}t_{\sf b}-t_{\sf ab}^2){\bf T}_i\cdot {\bf T}_j \right.  \right. \nonumber \\
& \quad +4t_{\sf ab}^2 T_i^{\sf x}T_j^{\sf x}
-4 (t_{\sf a}t_{\sf b} - t_{\sf ab}^2) T_i^{\sf y}T_j^{\sf y} 
 +(t_{\sf a}-t_{\sf b})^2 T_i^{\sf z}T_j^{\sf z}  \nonumber \\
& \quad +2 \sigma_{ij} t_{\sf ab}(t_{\sf a}-t_{\sf b})(T_i^{\sf x}T_j^{\sf z}+T_i^{\sf z}T_j^{\sf x})
+\frac{1}{2}(t_{\sf a}^2-t_{\sf b}^2)(T_i^{\sf z}+T_j^{\sf z}) \nonumber \\
& \quad \left. + \sigma_{ij} t_{\sf ab}(t_{\sf a}+t_{\sf b})(T_i^{\sf x}+T_j^{\sf x})
+\frac{1}{4}(t_{\sf a}^2+2t_{\sf ab}^2+t_{\sf b}^2) 
 \right] \nonumber \\
 &+\frac{1}{1-J/U}
 \mathcal{P}_{ij}^{\sf S=0}
 \left[
4 (t_{\sf a}t_{\sf b} - t_{\sf ab}^2) T_i^{\sf y}T_j^{\sf y} \right. \nonumber \\
& \quad +\frac{1}{2}(t_{\sf a}^2-t_{\sf b}^2)(T_i^{\sf z}+T_j^{\sf z}) 
+\sigma_{ij} t_{\sf ab}(t_{\sf a}+t_{\sf b})(T_i^{\sf x}+T_j^{\sf x}) \nonumber \\
& \quad \left. +\frac{1}{2}(t_{\sf a}^2+2t_{\sf ab}^2+t_{\sf b}^2) 
 \right] \nonumber \\
 &+\frac{1}{1-3J/U}
 \mathcal{P}_{ij}^{\sf S=1}
 \left[
-2(t_{\sf a}t_{\sf b}-t_{\sf ab}^2){\bf T}_i\cdot {\bf T}_j
-4t_{\sf ab}^2 T_i^{\sf x}T_j^{\sf x} \right. \nonumber \\
 & \quad -(t_{\sf a}-t_{\sf b})^2 T_i^{\sf z}T_j^{\sf z} 
 -2 \sigma_{ij} t_{\sf ab}(t_{\sf a}-t_{\sf b})(T_i^{\sf x}T_j^{\sf z}+T_i^{\sf z}T_j^{\sf x}) \nonumber \\
& \quad \left. \left. +\frac{1}{4}(t_{\sf a}^2+2t_{\sf ab}^2+t_{\sf b}^2) 
 \right] 
 \right\}.
 \label{eq:H-CuCuprime}
\end{align}
Here ${\sum_{ij}}^\prime$ denotes a sum over all ${\sf B}^\prime$ and ${\sf C}^\prime$ bonds, \mbox{$\sigma_{ij}=-1$} for ${\sf B}^\prime$ bonds and $\sigma_{ij}=1$ for ${\sf C}^\prime$ bonds (see Fig.~\ref{fig:dechoney} for bond labelling).
In consequence there are now 5 independent parameters: $t/t^\prime$, $t_{\sf a}/t^\prime$, $t_{\sf b}/t^\prime$, $t_{\sf ab}/t^\prime$ and $J/U$.
%


\subsection{Estimates of hopping parameters for Ba$_3$CuSb$_2$O$_9$}


Before mapping out the phase diagram on the decorated honeycomb lattice, it is useful to estimate the value of the hopping amplitudes in Ba$_3$CuSb$_2$O$_9$ in order to reduce the number of variable parameters.
Here we make rough estimates of $t/t^\prime$, $t_{\sf a}/t^\prime$, $t_{\sf b}/t^\prime$ and $t_{\sf ab}/t^\prime$, using tablulated values for the interatomic matrix elements associated with Cu-O  and O-O bonds\cite{harrison,slater54}.
\begin{table}
\begin{center}
  \begin{tabular}{| c | c | c | c | c |  }
    \hline 
\multirow{2}{*}{Hopping Parameter}  &
\multirow{2}{*}{$t/t^\prime$}  & 
\multirow{2}{*}{$t_{\sf a}/t^\prime$} &
\multirow{2}{*}{$t_{\sf b}/t^\prime$} &
\multirow{2}{*}{$t_{\sf ab}/t^\prime$}   \\ 
&  &  &  & \\ \hline
\multirow{2}{*}{Estimated value}  & 
\multirow{2}{*}{-1/3}  & 
\multirow{2}{*}{2/3} &
\multirow{2}{*}{0} &
\multirow{2}{*}{$-1/\sqrt{3}$}   \\ 
&  &  &  & \\ \hline 
    \end{tabular}
\end{center} 
\caption{\footnotesize{
Estimated values for the hopping parameters on the decorated honeycomb lattice. 
}}
\label{tab:hoppingparameters}
\end{table}

First we consider superexchange along Cu-O-O-Cu path 1 (see Fig.~\ref{fig:Ba3CuSb2O9-structure}) associated with {\sf A} bonds (see Fig.~\ref{fig:dechoney}).
Ignoring small deviations from octahedral symmetry, one finds,
\begin{align}
t &= -\frac{1}{4} V_{\sf pd\sigma} (V_{\sf pp\sigma}-V_{\sf pp\pi}) V_{\sf pd\sigma} \nonumber \\
t^\prime &= \frac{3}{4} V_{\sf pd\sigma} (V_{\sf pp\sigma}-V_{\sf pp\pi}) V_{\sf pd\sigma},
\end{align}
where, for example, $V_{\sf pd\sigma}$ is the interatomic matrix element for hopping between $\sigma$-bonded p and d orbitals.
This leads to $t/t^\prime = -1/3$.

Hopping along the Cu-O-O-Cu superexchange path 2 (see Fig.~\ref{fig:Ba3CuSb2O9-structure})  is similar, but with a rotated geometry.
We find $t_{\sf a}/t^\prime=2/3$, $t_{\sf b}/t^\prime=0$ and $t_{\sf ab}/t^\prime=-1/\sqrt{3}$.
These values are collected in Table~\ref{tab:hoppingparameters}.
%


\subsection{Ground state phase diagram for the 12-site cluster}


The ground state phase diagram on the decorated honeycomb lattice is calculated as a function of $t/t^\prime$ and $J/U$ using the parameters $t_{\sf a}/t^\prime=2/3$, $t_{\sf b}/t^\prime=0$ and $t_{\sf ab}/t^\prime=-1/\sqrt{3}$.
It is expected that Ba$_3$CuSb$_2$O$_9$ sits approximately at $t/t^\prime=-1/3$ and small $J/U$.
The phase diagram for a 12-site cluster with periodic boundary conditions, calculated both with Lanczos diagonalisation and spin-orbital decoupling, is shown in Fig.~\ref{fig:dechoneycomb-n12-pd}.

\begin{figure}[ht]
\centering
\includegraphics[width=0.3\textwidth]{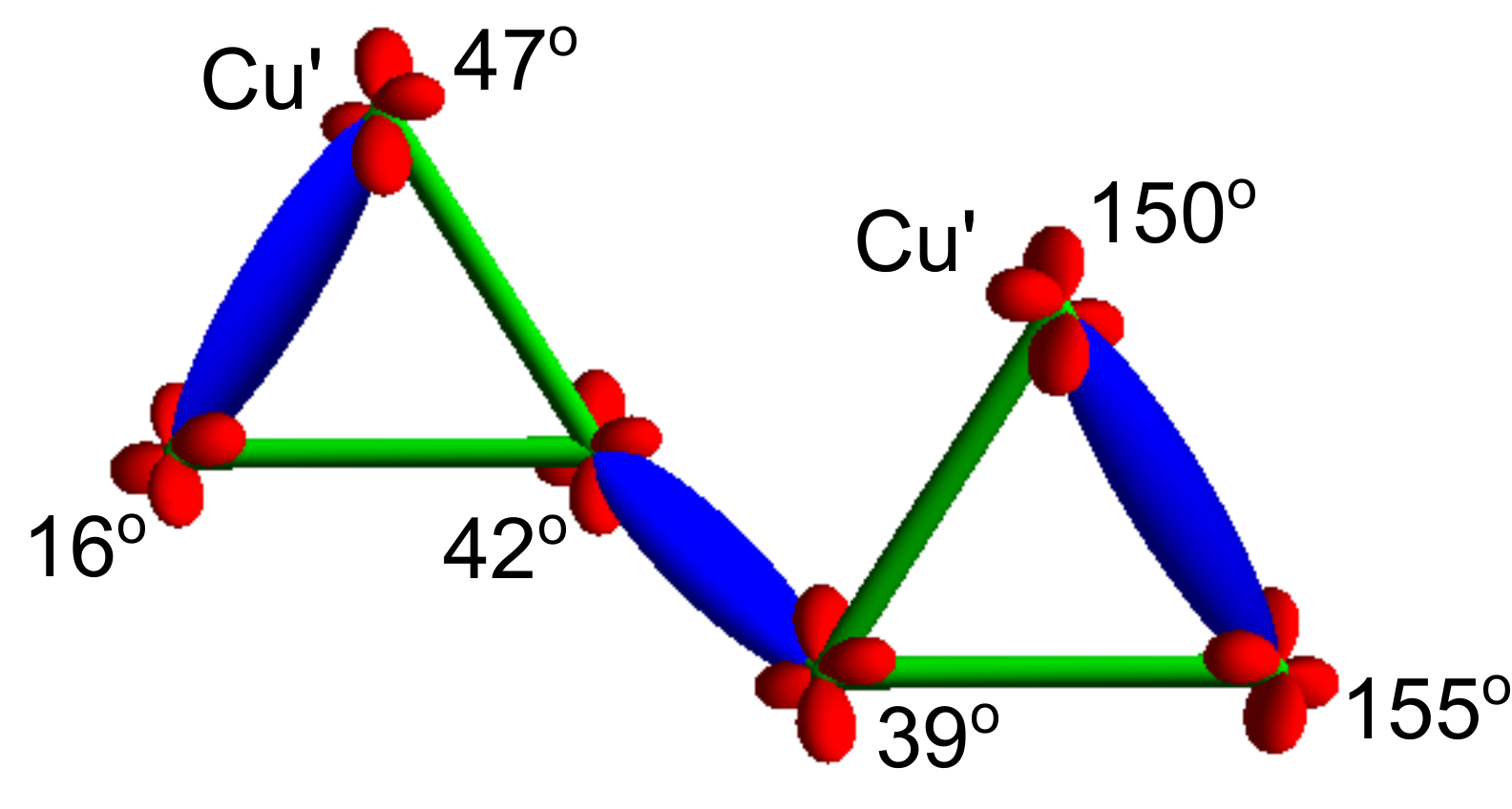}
\caption{\footnotesize{
6-site cluster composed of 4 Cu and 2 Cu$^\prime$ sites, useful for understanding the ground state of \mbox{$\mathcal{H}_{\sf ST} + \mathcal{H}_{\sf ST}^\prime$} [Eqs.~(\ref{eq:HST-honeycomb},\ref{eq:H-CuCuprime})] at small $J/U$.
Spins form nearest-neighbour singlets (shown in blue) on the 6-site cluster.
An angle \mbox{$\theta = \arctan [\langle T^{\sf z} \rangle/\langle T^{\sf x} \rangle]$} is used to specify the orbital degree of freedom (shown in red).
Angles are calculated for $t/t^\prime=-1/3$ and $J/U=0$, but are representative of the entire phase. 
}}
\label{fig:6sitedimer}
\end{figure}
\begin{figure}[ht]
\centering
\includegraphics[width=0.35\textwidth]{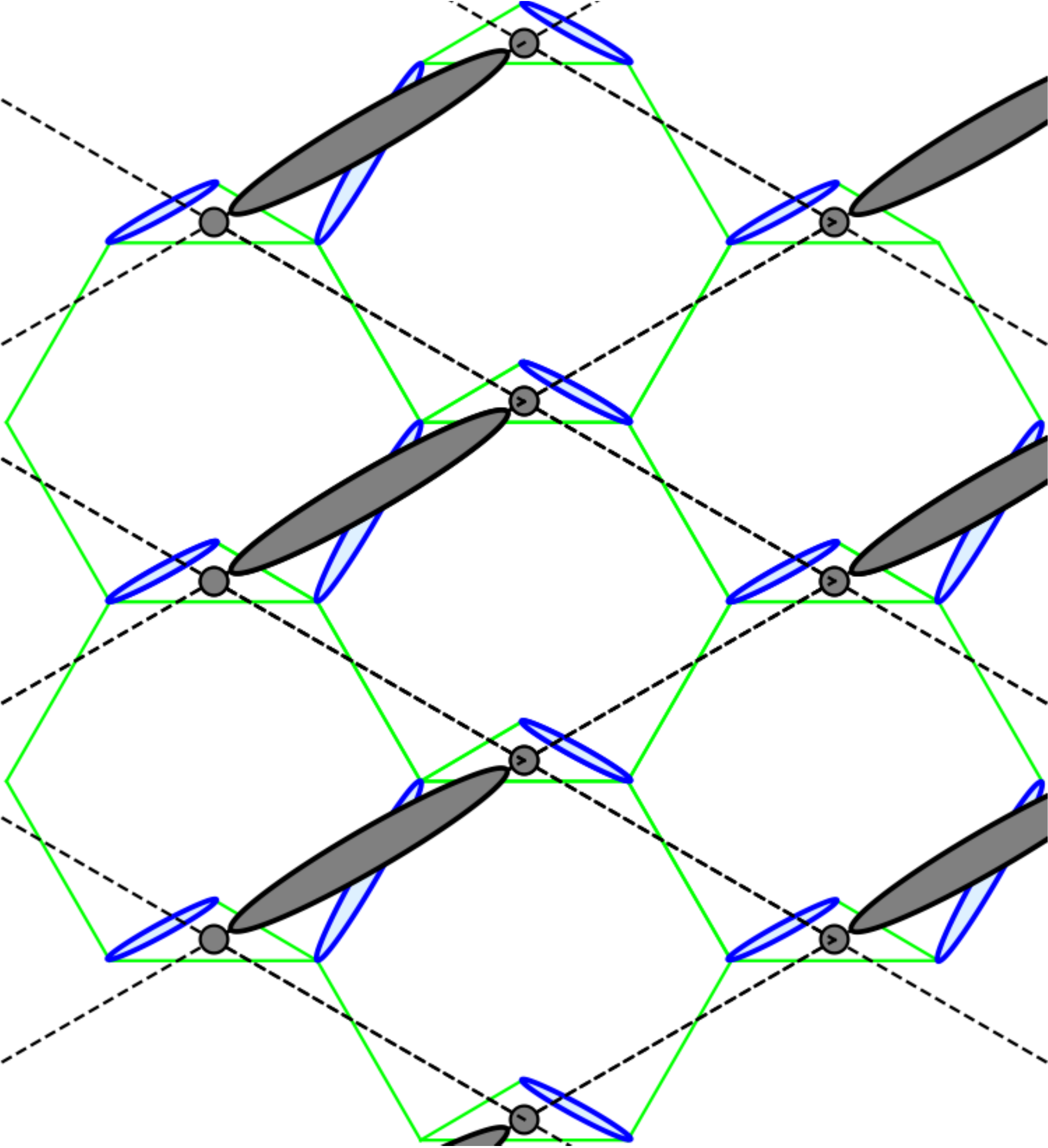}
\caption{\footnotesize{
Emergent square lattice useful for understanding the ground state of \mbox{$\mathcal{H}_{\sf ST} + \mathcal{H}_{\sf ST}^\prime$} [Eqs.~(\ref{eq:HST-honeycomb},\ref{eq:H-CuCuprime})] at small $J/U$.
The decorated honeycomb lattice (see Fig.~\ref{fig:dechoney}) is shown projected onto a plane (green).
Spin singlets (blue ellipses) form on this lattice in groups of three, with associated orbital state, as shown in Fig.~\ref{fig:6sitedimer}.
These can be thought of as dimers (grey ellipses) on an emergent square lattice (grey sites, black dashed bonds).
How the dimers are arranged on this lattice remains an open question.
}}
\label{fig:effsqlat}
\end{figure}

The most interesting phase occurs at small $J/U$ and is labelled as the emergent dimer phase (ED) in Fig.~\ref{fig:dechoneycomb-n12-pd}.
In this phase every Cu and Cu$^\prime$ spin forms a spin singlet with one of its nearest neighbours.
The basic unit is a 6-site cluster of four Cu sites and two Cu$^\prime$ sites, shown in Fig.~\ref{fig:6sitedimer}.
Within this cluster three nearest-neighbour singlet bonds form, two on Cu-Cu$^\prime$ bonds and one on a Cu-Cu bond.
The associated orbital configuration is shown in Fig.~\ref{fig:6sitedimer}.
These 6-site clusters can be thought of as a dimer on an emergent square lattice, as shown in Fig.~\ref{fig:effsqlat}.

How these dimers are arranged on the square lattice remains an open question.
To answer this question within a numerical diagonalisation approach would require significantly larger cluster sizes.
Another way this issue could be resolved would be to derive an effective quantum dimer Hamiltonian on the emergent square lattice.
Dynamical processes would require breaking at least 4 spin-singlet bonds, and therefore would only enter at high order in perturbation theory.

\begin{figure}[ht]
\centering
\includegraphics[width=0.45\textwidth]{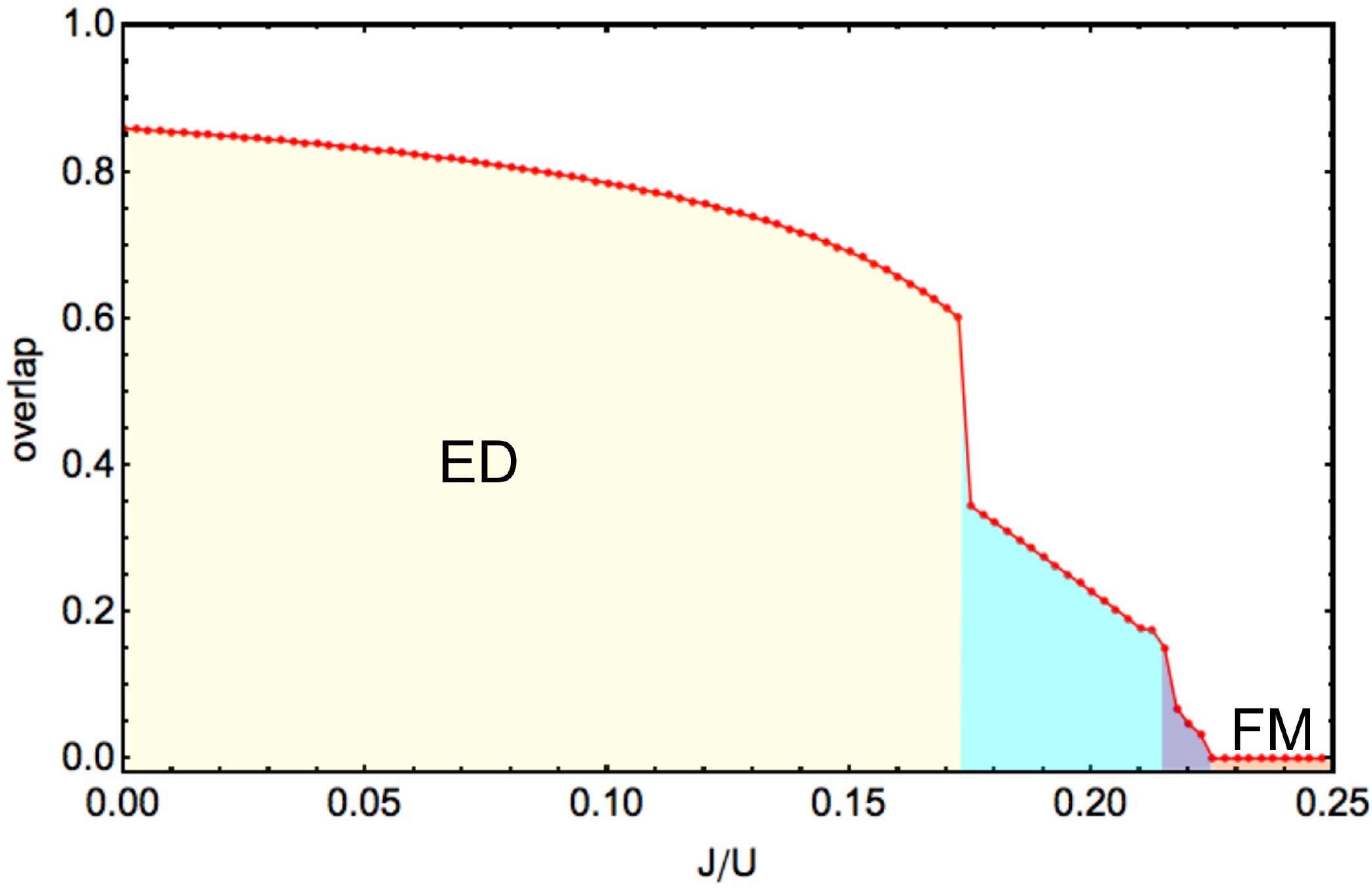}
\caption{\footnotesize{
Overlap between the ground state wavefunction determined by lanczos diagonalisation and a linear superposition of the 8 orientations of the  emergent dimer (ED) state found using the spin-orbital decoupling scheme.
The coefficients of the superposition are adjusted to maximise the overlap.
The overlap is plotted as a function of $J/U$, with $t/t^\prime=-1/3$.
The overlap is significant throughout the ED phase.  
}}
\label{fig:overlap}
\end{figure}

In order to confirm that the same ED phase is found with both the spin-orbital decoupling and the Lanczos diagonalisation methods, we calculated the overlap between the Lanczos wavefunction and a superposition of the 8 dimer coverings possible on the emergent 4-site square lattice.
These dimer covering wavefunctions are computed within the spin-orbital decoupling scheme.
The overlap is shown in Fig.~\ref{fig:overlap}, and is significant throughout the ED phase.

The phase diagram, Fig.~\ref{fig:dechoneycomb-n12-pd}, includes a number of other phases.
Unlike on the honeycomb lattice, the point \mbox{$t/t^\prime=1$}, $J/U=0$ is no longer {\sf SU(4)} symmetric, due to the choice of $t_{\sf a}/t^\prime$, $t_{\sf b}/t^\prime$ and $t_{\sf ab}/t^\prime$.
Instead it is part of a spin singlet phase (coloured red) with the same spin configuration as the ED phase but different orbital configuration.

The pair of $S=2$ phases (coloured green) have spin ferromagnetism on the honeycomb lattice with opposite spin direction on the Cu$^\prime$ sites.
Orbitals are $d^{\sf x^2-y^2}$-like on the honeycomb lattice and $d^{\sf 3z^2-r^2}$-like on the Cu$^\prime$ sites.

At large $J/U$ the spins are ferromagnetic (coloured orange, labelled FM).
One way to gain some understanding of the orbital state is to assume large Cu-Cu$^\prime$ hopping amplitudes, and to perform perturbation theory with $t$ and $t^\prime$ as small numbers.
Numerically on a 12-site cluster there is no obvious phase transition as the Cu-Cu$^\prime$ hopping amplitudes are scaled back to physical values.
The suggestion is that the orbitals form an incommensurate order with variable ordering vector.

In addition, there are some intermediate phases (coloured blue) between the ED phase and the FM phase.
These do not match between the Lanczos diagonalisation and the spin-orbital decoupling.
In the spin orbital decoupling phase diagram, the dark blue $S=0$ phase has ferromagnetic chains aligning antiferromagnetically with neighbouring chains.
The wavefunction of this state has a significant overlap with the 1st excited state of the $S=4$ phase found in Lanczos diagonalisation (coloured dark blue).

Finally, it is interesting to study the crossover between the honeycomb lattice phase diagram (Fig.~\ref{fig:honeycomb-n6-pd} and Fig.~\ref{fig:honeycomb-n18-pd}) and the decorated honeycomb lattice (Fig.~\ref{fig:dechoneycomb-n12-pd}).
We consider the point $t/t^\prime=-1/3$, $J/U=0$ and parametrise the Cu-Cu$^\prime$ hopping as $t_{\sf a}/t^\prime = 2\tilde{t}/3$, $t_{\sf b}/t^\prime = 0$ and $t_{\sf ab}/t^\prime = -\tilde{t}/\sqrt{3}$, that is the ratio $t_{\sf a}:t_{\sf b}:t_{\sf ab}$ is kept fixed, but the magnitude is varied.
Using the spin-orbital decoupling method with a 12-site cluster, we find that the collinear dimer state (CD) is preferred in the region $\tilde{t} \lesssim 0.5$, while for $\tilde{t} \gtrsim 0.7$ the ED phase is stable.


\section{Discussion and conclusions}
\label{sec:discussion}


Finally, we discuss the experimental situation in Ba$_3$CuSb$_2$O$_9$, keeping in mind the results of Section~\ref{sec:honeycomb} and Section~\ref{sec:dechoneycomb}.
Realistic parameters for Ba$_3$CuSb$_2$O$_9$ are $t/t^\prime \approx -1/3$, $J/U\approx 0$, and therefore we concentrate in particular on the CD phase found on the honeycomb lattice (Fig.~\ref{fig:honeycomb-n18-pd}) and the ED phase found on the decorated honeycomb lattice (Fig.~\ref{fig:dechoneycomb-n12-pd}).

Interpretation of the experimental data is complicated by the structural disorder, which originates from the Ising choice associated with the dumbbell orientation.
However, since both the ED and the CD phases are based on nearest-neighbour spin-singlet bonds, and are therefore local in nature, they should provide a good description of the nanoscale domains found in Ba$_3$CuSb$_2$O$_9$.
Measurements show that the honeycomb lattice of Cu ions has a structural correlation length of $\sim$10\AA, corresponding to about twice the Cu-Cu inter-ion spacing\cite{nakatsuji12}.
Small regions of the material, consisting of approximately 6-20 Cu ions, can be thought of in terms of the clusters studied in Section~\ref{sec:honeycomb} and Section~\ref{sec:dechoneycomb}, albeit with more complicated boundary conditions.
For most of these small regions one should consider the decorated honeycomb lattice, but there will also be small regions in which the physics of the honeycomb lattice is relevant.
Inelastic neutron scattering\cite{nakatsuji12} and NMR $1/T_1$ relaxation\cite{quilliam12} studies show that the majority of the spins form singlet bonds, and they see evidence for a singlet-triplet excitation gap of roughly 50K.
This is consistent with both the CD and ED phases (see Fig.~\ref{fig:honeycomb-n18-pd} and Fig.~\ref{fig:dechoneycomb-n12-pd}).
Fits to neutron data for the equal-time correlation function extract a characteristic spatial separation for the singlet bonds of 5.6(1)\AA\cite{nakatsuji12}. 
It is intriguing to notice that this number is closer to the Cu-Cu$^\prime$ bond distance of 5.61\AA \ than the Cu-Cu bond distance of 5.81\AA.
This provides tentative support to the existence of the ED phase, where the ratio of Cu-Cu$^\prime$ to Cu-Cu bonds is 2:1.
It would be interesting if the equal time correlation function could be resolved into two components, one with characteristic length of 5.61\AA \ and the other at  5.81\AA.

Another interesting possibility is the resolution of the singlet-triplet excitation into two distinct energy gaps.
The basic unit of the ED phase is the 6-site cluster shown in Fig.~\ref{fig:6sitedimer}.
Using the spin-orbital decoupling approach, it is possible to compare the spin-singlet ground state with excited states in which one of the three singlet bonds has been promoted to a triplet.
One finds that the singlet-triplet excitation should be resolvable into two components:  one associated with one of the Cu-Cu$^\prime$ bonds, and a second at $\sim$1.25 times the energy associated with the Cu-Cu bond and the other Cu-Cu$^\prime$ bond.
In the ED phase the weight associated with these excitations should be in the ratio 1:2.
In reality, the structural disorder is going to considerably broaden these excitations, but they may still be resolvable.

Not all the spins form singlets, and there is a sizeable minority of weakly interacting spins.
Magnetisation measurements show that $\sim$16\% of the spins are ``orphaned''\cite{quilliam12}.
It has been suggested that these are associated with the Cu$^\prime$ sites\cite{quilliam12}.
However, for a long-range ordered decorated honeycomb lattice, 33\% of the sites are Cu$^\prime$ and it is difficult to reconcile this with the measured 16\% of spins weakly interacting.
Instead we suggest that the orphan spins occupy both Cu and Cu$^\prime$ sites, and arise due to geometric constraints associated with the structural disorder.
This has support from electron spin resonance (ESR) measurements, which show an isotropic response, consistent with the idea that the weakly interacting spins are distributed over multiple sites\cite{nakatsuji12}.

In order to determine the orbital state, a number of structural measurements have been made, probing the nature of the Jahn-Teller distortions.
X-ray diffraction studies of non-stoichiometric Ba$_3$CuSb$_2$O$_9$ samples show a long-range orthorhombic distortion, with four short and two long Cu-O bonds.
This shows that the holes occupy $d^{\sf x^2-y^2}$-type orbitals.
These measurements are consistent with the CD phase that we find on the honeycomb lattice (see Fig.~\ref{fig:honeycomb-n18-pd}).

Stoichiometric samples are more complicated.
Extended x-ray absorption fine structure (EXAFS) studies, which probe at timescales of $10^{-16}$s, see no difference between the orthorhombically distorted non-stoichiometric samples and the stoichiometric samples\cite{nakatsuji12} at 10K.
However, x-ray diffraction\cite{nakatsuji12,katayama14}, ESR absorption\cite{nakatsuji12} and Raman spectroscopy\cite{katayama14}, which probe on longer timescales, see a hexagonally symmetric crystal, with no evidence for an orthorhombic distortion.
One proposed explanation is that the orbitals fluctuate on a timescale intermediate between the $10^{-16}$s of the EXAFS measurements and the approximately $10^{-11}$s timescale of ESR\cite{nakatsuji12,nasu13,ishiguro13,katayama14}.
Another possibility is that the system undergoes a static short-range distortion, and the ESR, x-ray and Raman experiments probe clusters with large enough size that the spatial distortions average out, restoring hexagonal symmetry\cite{nakatsuji12,quilliam12}.

The theory presented here does not provide a definitive answer to the question of whether there is a fluctuating Jahn-Teller distortion.
However, it is interesting to speculate on what mechanism could drive an orbital fluctuation.
The structural disorder splits the lattice into nanoscale clusters.
A nanoscale cluster of Cu ions will be completely surrounded by Sb ions, and will therefore only interact very weakly with rest of the system.
If one attempts to cover each nanocluster with as many nearest-neighbour spin-singlet bonds as possible, consistent with the ED phase found in Section~\ref{sec:dechoneycomb}, there will in general be multiple coverings.
These dimer coverings of the nanoscale domains describe a low energy subspace, and at low temperatures one would expect resonance between the different spin-singlet configurations.
Furthermore, spin resonance would be accompanied by orbital resonance, driving a fluctuating Jahn-Teller distortion.
The driving force could be some combination of spin-orbital exchange via \mbox{$\mathcal{H}_{\sf ST} + \mathcal{H}_{\sf ST}^\prime$} [Eqs.~(\ref{eq:HST-honeycomb},\ref{eq:H-CuCuprime})] and vibronic tunnelling effects, of the type considered in Ref.~[\onlinecite{nasu13}]. 
This would be one way to account for the isotropic signal seen for example in ESR studies\cite{nakatsuji12}.
We think this would be an interesting avenue to explore in future studies.

Finally we briefly mention measurements of diffuse x-ray scattering\cite{ishiguro13}.
The diffuse intensity surrounding the 220 Bragg peak shows a high intensity region of scattering surrounding $(\Delta,\delta)=(2,\pm0.03)$ (see Ref.~[\onlinecite{ishiguro13}] for more details).
This can be modelled relatively easily, for example by considering small clusters realising either the CD phase (see Fig.~\ref{fig:honeycomb-n18-pd}) or the ED phase (see Fig.~\ref{fig:dechoneycomb-n12-pd}).
Conversely, lobes of scattering around $(\Delta,\delta)=(2\pm0.05,\pm0.05)$ are far more difficult to model.
They cannot be reconciled with either the spin-orbital resonant state suggested in Ref.~[\onlinecite{nasu13}] (which is related to the NCD state found for positive $t/t^\prime$ in Fig.~\ref{fig:honeycomb-n18-pd}), or with the ED state shown in Fig.~\ref{fig:dechoneycomb-n12-pd}.
The only simple way we have found to model these lobes on small clusters is by considering antiferro-orbital bonds, where one site has a $d^{\sf x^2-y^2}$-type orbital and the other a $d^{\sf 3z^2-r^2}$-type orbital.
The place we find such an orbital phase in the above theory is in the $S=2$ phase on the decorated honeycomb lattice (see Fig.~\ref{fig:dechoneycomb-n12-pd})  or in the FM phase on the honeycomb lattice (see Fig.~\ref{fig:honeycomb-n18-pd}).
While modelling the diffuse x-ray scattering clearly requires a superposition of clusters with different orbital states, a full explanation of the data remains an interesting open question, and in need of further calculations.

In conclusion, we have studied a spin-orbital model relevant to Ba$_3$CuSb$_2$O$_9$.
This was derived from a Hubbard model, which has its origins in the quantum chemistry of the material.
We have determined the phase diagram on both the honeycomb and the decorated honeycomb lattices by considering small clusters, discovering a rich array of phases.
While the 2-dimensional honeycomb lattice is not directly relevant to Ba$_3$CuSb$_2$O$_9$, we have shown that an {\sf SU(4)} spin-orbital liquid phase exists over a wide range of parameters.
We hope this will motivate future attempts to synthesise honeycomb lattice materials with active spin and orbital degrees of freedom.
On the decorated honeycomb lattice, and for parameters relevant to Ba$_3$CuSb$_2$O$_9$, we have found a phase dubbed the emergent dimer (ED) phase, which involves nearest-neighbour spin singlets, orbital order and can be thought of as a set of dimers on an emergent square lattice.
When one considers that Ba$_3$CuSb$_2$O$_9$ in fact consists of nanoscale clusters with only short-range decorated honeycomb order, the ED phase can be thought of as defining a low energy subspace for the clusters.
These findings are consistent with a range of experimental measurements.


{\it Acknowledgments.}   
We are grateful to Fran\c cois Vernay for useful discussions.
We thank the Swiss National Science Foundation and its SINERGIA network ``Mott physics beyond the Heisenberg model'' for financial support.


\appendix



\section{Spin-orbital decoupling approximation}
\label{App:SOdecoup}


Here we provide details of the spin-orbital decoupling mean field approximation, used to calculate ground state phase diagrams in Fig.~\ref{fig:honeycomb-n6-pd}, Fig.~\ref{fig:honeycomb-n18-pd} and Fig.~\ref{fig:dechoneycomb-n12-pd}.
The method provides a significant reduction in the size of the Hilbert space, as compared to full diagonalisation.
The technique was developed in Refs.~[\onlinecite{vernay04,mila98}].

Hamiltonians of the form $\mathcal{H}_{\sf ST}$ [Eq.~(\ref{eq:HST-honeycomb})] can be written compactly as,
\begin{align}
\mathcal{H}_{\sf ST} &= \frac{4t^2}{U} \sum_{\langle ij \rangle} 
\left\{
2({\bf S}_i\cdot {\bf S}_j ) {\bf h}^{\sf T}_{ij} + {\bf k}^{\sf T}_{ij} 
\right\} .
\end{align}
We make the ansatz that the spin and orbital dependence of the wavefunction can be decoupled as,
\begin{align}
| \Psi \rangle  = | \Psi^{\sf S} \rangle \otimes  | \Psi^{\sf T} \rangle.
\label{eq:ansatz}
\end{align}
Clearly this approximation breaks down close to the {\sf SU(4)} point, where spin and orbital degrees of freedom are intimately coupled.
However, elsewhere it is expected to work well, and Fig.~\ref{fig:honeycomb-n6-pd} shows it can produce very similar results to exact diagonalisation.
Using the decoupling ansatz, Eq.~\ref{eq:ansatz}, one can write two Hamiltonians, one averaged over $| \Psi^{\sf S} \rangle$,
\begin{align}
\mathcal{H}_{\sf T} &=
\langle \Psi^{\sf S} | \mathcal{H}_{\sf ST} | \Psi^{\sf S} \rangle \nonumber \\
&= \frac{4t^2}{U} \sum_{\langle ij \rangle} 
\left\{
\langle \Psi^{\sf S} | 2{\bf S}_i\cdot {\bf S}_j | \Psi^{\sf S} \rangle {\bf h}^{\sf T}_{ij} + {\bf k}^{\sf T}_{ij} 
\right\} ,
\label{eq:HT}
\end{align}
and the other averaged over $| \Psi^{\sf T} \rangle$,
\begin{align}
\mathcal{H}_{\sf S} &=
\langle \Psi^{\sf T} | \mathcal{H}_{\sf ST} | \Psi^{\sf T} \rangle \nonumber \\
&= \frac{4t^2}{U} \sum_{\langle ij \rangle} 
\left\{
2 {\bf S}_i\cdot {\bf S}_j   \langle \Psi^{\sf T} | {\bf h}^{\sf T}_{ij} | \Psi^{\sf T} \rangle 
+ \langle \Psi^{\sf T} | {\bf k}^{\sf T}_{ij} | \Psi^{\sf T} \rangle
\right\} .
\label{eq:HS}
\end{align}
In order to find the ground state within this mean-field approximation, it is necessary to solve the equations,
\begin{align}
&\mathcal{H}_{\sf S}  | \Psi^{\sf S} \rangle = E_{\sf S} | \Psi^{\sf S} \rangle \nonumber \\
&\mathcal{H}_{\sf T}  | \Psi^{\sf T} \rangle = E_{\sf T} | \Psi^{\sf T} \rangle,
\end{align}
self consistently.
This can be done by guessing a starting spin configuration, diagonalising $\mathcal{H}_{\sf T}$ [Eq.~\ref{eq:HT}] to find the orbital configuration, feeding this into $\mathcal{H}_{\sf S}$ [Eq.~\ref{eq:HS}], diagonalising to find the spin configuration, and looping until the ground-state eigenvalue is self consistent.

The main challenge with this method is that it is common to converge to a local minima rather than a global minima.
To overcome this problem one can either increase the number of eigenvalues retained at each step or one can consider many different, randomly chosen, starting configurations.
In practice it is useful to use a combination of these two strategies.

\bibliographystyle{apsrev4-1}
\bibliography{bibfile}

\end{document}